\documentclass[11pt,a4paper]{article}
\pdfoutput=1

\usepackage{jheppub}
\usepackage{rotating}

\usepackage{wrapfig}
\usepackage{orcidlink}
\usepackage{bbm}
\usepackage{soul}
\usepackage{amsfonts}
\usepackage{amsmath}
\usepackage{slashed}
\usepackage{amssymb}
\usepackage{latexsym}
\usepackage{multirow}
\usepackage{hyperref}
\usepackage{enumitem}
\hypersetup{colorlinks=true,citecolor=red,linkcolor=magenta,urlcolor=blue}
\usepackage{subfigure}
\usepackage{relsize}
\usepackage{cleveref}
\usepackage{fancyhdr}
\usepackage{float}
\usepackage{graphicx}
\usepackage{microtype}
\usepackage{tabularx}
\usepackage{xcolor}
\usepackage{tensor}
\usepackage[bottom]{footmisc}
\usepackage{tikz}
\usepackage{tikz-feynman}
\usetikzlibrary{arrows,arrows.meta,shapes,positioning,calc,backgrounds}
\usetikzlibrary{fadings,shapes.arrows,shadows}
\usetikzlibrary{decorations.markings}
\usetikzlibrary{decorations.pathmorphing}
\usetikzlibrary{decorations.text}
\usetikzlibrary{shapes.misc,shapes.geometric}

\newcolumntype{C}{>{$}c<{$}}
\newcolumntype{L}{>{$}l<{$}}
\newcolumntype{R}{>{$}r<{$}}

\newcommand{\mg}{\textsc{MadGraph5\_}aMC@NLO}

\title{Doubling down on down-type diquarks}
\author[a]{Christoph Englert\orcidlink{0000-0003-2201-0667},}
\emailAdd{christoph.englert@glasgow.ac.uk}
\author[a,b]{Christiane Mayer\orcidlink{0009-0007-6692-4977},}
\emailAdd{christiane.mayer@physik.uzh.ch}
\author[a]{Wrishik Naskar\orcidlink{0000-0002-4357-8991},}
\emailAdd{w.naskar.1@research.gla.ac.uk}
\author[a]{Sophie Renner\orcidlink{0000-0002-8374-769X}}
\emailAdd{sophie.renner@glasgow.ac.uk}

\affiliation[a]{School of Physics \& Astronomy, University of Glasgow, Glasgow G12 8QQ, United Kingdom}
\affiliation[b]{Physik-Institut, Universit\"at Z\"urich, CH-8057 Z\"urich, Switzerland}

\abstract{Effective field theory-based searches for new physics at colliders are relatively insensitive to interactions involving only right-handed down-type quarks. These interactions can hide amongst jet backgrounds at the LHC, and their indirect effects in electroweak and Higgs processes are small. Identifying scenarios in which these interactions dominate, we can naturally pick out just two tree-level mediators, both scalar diquarks. Over the full parameter space of these states, we analyse exotics searches at current and future hadron colliders, Higgs signal strength constraints, and indirect constraints from flavour physics, finding genuine complementarity between the data sets. In particular, while flavour constraints can exclude diquarks in the hundreds of TeV mass range, these can be evaded once a flavour structure is imposed on the couplings, as we illustrate by embedding the diquarks within a composite Higgs model. In combination, however, we show that flavour and collider constraints exclude down-type diquarks to multi-TeV scales, thus narrowing the remaining hiding places for new interactions amongst LHC data.}

\begin{document}
\maketitle
\allowdisplaybreaks
\flushbottom
\section{Introduction}
\label{sec:intro}
 As the Large Hadron Collider (LHC) now operates close to its designed centre-of-mass energy, direct search limits on exotic BSM states with non-trivial Quantum Chromodynamics (QCD) quantum numbers will explore new parameter regions of new physics scenarios. The experiments still have significant potential to improve upon traditional cut-and-count analyses by relying on a fast-developing arsenal of multivariate and machine-learning techniques. These have proved very successful in removing background pollution and reducing systematics. To inform these ongoing efforts, it appears to be timely to revisit the relevance of such searches in light of available precision flavour measurements. This clarifies the achievable new physics (NP) discovery potential at the high-luminosity (HL) LHC, specifically by combining the precision and energy frontiers using well-motivated new physics scenarios that leave footprints in both phenomenological arenas. Especially motivated in this regard are models that introduce new sources of flavour violation in the multi-TeV range.

Under the assumption of a reasonably large mass gap between the new physics and the Standard Model (which appears motivated given the current null results of the LHC's exotics programme), the Standard Model Effective Field Theory (SMEFT)~\cite{Grzadkowski:2010es} appears as a motivated approach to compare exotics searches at the LHC with flavour observables. Although many SMEFT interactions are well explored and constrained in the literature, a notable exception is the operator containing only right-handed down-type quarks
\begin{equation}
O_{dd}= (\bar{d}^i_R\gamma^\mu d^j_R)(\bar{d}^k_R\gamma_\mu d^l_R).
\end{equation}
Traditional tree-level SMEFT fits (e.g.~\cite{Ellis:2020unq,Elmer:2023wtr,Celada:2024mcf}) do not constrain this operator at all, since it does not contribute to electroweak, Higgs or top observables at leading order. Including renormalisation-group evolution, $O_{dd}$ can run into electroweak precision and top observables, but the resulting constraints remain weak, $\Lambda \sim 0.1$ -- $0.4~\text{TeV}$ for the effective NP scale~\cite{Allwicher:2023shc,Greljo:2023bdy}. The only collider observables that the $O_{dd}$ operator enters at tree level are LHC dijet distributions, but this operator has not as yet been individually constrained in dijet fits~\cite{CMS:2018ucw,ATLAS:2017eqx}. Furthermore, non-resonant analyses are challenging as they require a good understanding of large QCD jet backgrounds, which heavily relies on data-driven methods with fewer phenomenological handles than resonance production. Therefore, non-resonant analyses typically exhibit a reduced future performance increase compared to resonance searches.

Given the low indirect bounds on $O_{dd}$, another angle of attack on these operators is to study bounds on UV states which generate them at tree level. There are only two BSM particles that match \emph{only} onto $O_{dd}$ at tree level (and are therefore not better constrained by other operator bounds), as defined by their charges:
\begin{equation}
\label{eq:new}
\begin{split}
\Phi_{(3)} &\sim ({\bf{3}},{\bf{1}})_{2/3} ,\\
\Phi_{(6)} &\sim (\bar {\bf{6}},{\bf{1}})_{2/3},
\end{split}
\end{equation}
under $SU(3)_C\times SU(2)_L \times U(1)_Y$~\cite{deBlas:2017xtg}. These are both scalar diquarks, labelled as `VII' and `VIII' in the notation of Ref.~\cite{Giudice:2011ak}, or as $\omega_2$ and $\Omega_2$ in the notation of Ref.~\cite{deBlas:2017xtg}. We will focus on these scenarios in this work. The colour representations of the diquarks impose different flavour structures on their couplings; the triplet diquark has a flavour-antisymmetric coupling to quarks, while the sextet diquark has a flavour-symmetric coupling. This gives the two states rather different flavour phenomenology and affects how they can be produced at colliders.  

The LHC phenomenology of (one or both) down-type diquarks has been studied previously in Refs~\cite{Atag:1998xq,Plehn:2000be,Han:2009ya,Han:2010rf,Giudice:2011ak,Pascual-Dias:2020hxo}, while Refs~\cite{Barr:1989fi,Giudice:2011ak,Chen:2018dfc} study their flavour phenomenology.\footnote{Studies of diquarks with different quantum numbers, and hence inducing different four-quark operators, have recently been undertaken in Refs~\cite{Crivellin:2023saq,Bordone:2021cca}, with the aim of explaining anomalous results in flavour physics.} Our work seeks to add some important new aspects to this literature. Notably, LHC data and searches have not yet been interpreted as limits on the down-type sextet diquark, since previous collider phenomenology studies of this model all date from over a decade ago.\footnote{Exceptions are the studies of Refs~\cite{Carpenter:2021rkl,Carpenter:2022qsw,Carpenter:2023aec}, which focus on additional non-renormalisable interactions of colour sextets, and hence do not directly apply to the minimal case considered here.} Likewise, the effects of the off-diagonal couplings of the sextet in flavour physics have not been previously studied, owing to the fact that the products of the diagonal couplings are extremely strongly constrained by meson mixing observables. By contrast, its off-diagonal couplings contribute either at loop level or to less precisely measured observables, so can be safely ignored if they are of the same order as the diagonal ones. However, flavour symmetries or paradigms can plausibly suppress the most flavour-violating parts of the coupling matrix or the resulting effective operator and upset this assumption (see e.g.~\cite{Greljo:2023adz}). To provide a broad survey of diquark phenomenology, we therefore detail the first calculations of some loop-induced diquark flavour processes and include observables in our study which are sensitive to as many different combinations of couplings as possible. This allows us to study the complementarity of collider and flavour sensitivity in many areas of the diquark parameter space.

The structure of the paper is as follows. Firstly, in Sec.~\ref{sec:down}, we review the interactions of the down-type diquark triplets and sextets. Integrating out these states, we also detail the matching to the SMEFT formalism below their respective mass scales. Additionally, we comment on possible UV completions informed by compositeness. In Sec.~\ref{sec:collider}, we recast existing searches to obtain constraints on the model parameter space from resonance searches at the LHC. In passing, we also comment on constraints that can be expected for interactions of these states with the SM Higgs sector, and on the future sensitivity during the HL phase of the LHC and at possible future hadron machines. Sec.~\ref{sec:flavour} is dedicated to constraints from flavour searches, highlighting the difference in the phenomenology of the different scalar diquarks of interest. Sec.~\ref{sec:flavourcoll} discusses the comparison as well as the complementarity of flavour and collider searches for the models of Eq.~\eqref{eq:new}. We conclude in Sec.~\ref{sec:conc}.

\section{Down-type diquarks}
\label{sec:down}
Having argued in the Introduction that the right-handed down-type operator $O_{dd}= (\bar{d}^i_R\gamma^\mu d^j_R)(\bar{d}^k_R\gamma_\mu d^l_R)$ is hard to access purely within the EFT, we begin this section by justifying our focus on two scalar diquarks as a reasonable proxy for a situation in which this operator dominates at low energies.

There are only four BSM states which match to the effective operator $O_{dd}$ at tree level~\cite{deBlas:2017xtg}; two scalars and two vectors, as listed in Tab.~\ref{tab:completions}. The vectors $\mathcal{B}$ and $\mathcal{G}$ are heavy copies of the $Z$ and the gluon, respectively, and hence in general match at tree level to many other operators besides $O_{dd}$. These produce a range of signals, for example, $\mathcal{B}$ generates tree-level contributions to the $T$ parameter and other electroweak precision observables. $\mathcal{G}$ generates operators which contribute to top quark pair production at tree level. For these vector states, then, unless many of their couplings are set to zero, these other operators would likely show up before $O_{dd}$.\footnote{We note however that charging these states as non-trivial irreps of the down-type flavour symmetry group can provide a mechanism to forbid all other operators~\cite{Greljo:2023adz}, while inevitably implying flavour symmetry of the $O_{dd}$ operator itself. We do not explore this option here.} By contrast, the scalars $\Phi_{(3)}$ and $\Phi_{(6)}$ are forbidden by SM gauge symmetry from generating any other operators besides $O_{dd}$ at tree level. These two states, therefore, provide the most natural completion for a SMEFT in which $O_{dd}$ dominates.

\begin{table}
\begin{center}
\begin{tabular}{c| c| c | l}
State & Spin & SM charges & Tree-level generated operators \\
\hline
$\Phi_{(3)}$ & 0 & $({\bf{3}},{\bf{1}})_{2/3}$ & $O_{dd}$\\
$\Phi_{(6)}$ & 0 & $(\bar {\bf{6}},{\bf{1}})_{2/3}$ & $O_{dd}$\\
$\mathcal{B}$ & 1 & $({\bf{1}},{\bf{1}})_{0}$ & $O_{ll}$, $O_{qq}^{(1)}$, $O_{lq}^{(1)}$, $O_{ee}$, $O_{dd}$, $O_{uu}$, $O_{ud}^{(1)}$, $O_{le}$, $O_{qe}$, $O_{ld}$, $O_{lu}$, $O_{qd}^{(1)}$,\\
 &  & & $O_{qu}^{(1)}$, $O_{HD}$, $O_{H\Box}$, $O_{eH}$, $O_{dH}$, $O_{uH}$, $O_{Hl}^{(1)}$, $O_{Hq}^{(1)}$, $O_{He}$, $O_{Hd}$, $O_{Hu}$\\
$\mathcal{G}$ & 1 & $({\bf{8}},{\bf{1}})_{0}$ & $O_{qq}^{(1)}$, $O_{qq}^{(3)}$, $O_{dd}$, $O_{uu}$, $O_{ud}^{(8)}$, $O_{qu}^{(8)}$, $O_{qd}^{(8)}$\\
\end{tabular}
\caption{Complete list of single-particle UV completions which can generate the $O_{dd}$ operator at tree level, along with any other operators that are also tree-level generated by the same state. Taken from Ref.~\cite{deBlas:2017xtg}.\label{tab:completions}}
\end{center}
\end{table}

The Lagrangians for each diquark are
\begin{align}
\label{eq:conventions}
\mathcal{L}_{(6) }&= - y_{ij}^{(6)} \Phi_{(6)}^{(ab)}\,d_{Ri}^{Ta} C d_{Rj}^b+ {\text{h.c.}},\\
\mathcal{L}_{(3) }&=-y_{ij}^{(3)} \Phi_{(3)}^{a}\,\epsilon_{abc} \,d_{Ri}^{T b} C d_{Rj}^{c}+ {\text{h.c.}},
\end{align}
where $i,j$ are flavour indices, $a,b,c$ are fundamental colour indices, and $C$ is the charge conjugation matrix. The sextet coupling $y_{ij}^{(6)}$ is a symmetric complex matrix, while the triplet coupling  $y_{ij}^{(3)}$ is an antisymmetric complex matrix
\begin{equation}
\label{eq:couplingmatrices}
y^{(6)} = \begin{pmatrix}
y_{11}^{(6)} e^{i\phi_{11}} & y_{12}^{(6)} e^{i\phi_{12}} &  y_{13}^{(6)} e^{i\phi_{13}} \\
y_{12}^{(6)} e^{i\phi_{12}} & y_{22}^{(6)} e^{i\phi_{22}} &  y_{23}^{(6)} e^{i\phi_{23}} \\
y_{13}^{(6)} e^{i\phi_{13}} & y_{23}^{(6)} e^{i\phi_{23}} &  y_{33}^{(6)} e^{i\phi_{33}} \\
\end{pmatrix}, ~~~~y^{(3)} = \begin{pmatrix}
0 & y^{(3)}_{12} e^{i\delta_{12}} &  y^{(3)}_{13} e^{i\delta_{13}} \\
-y^{(3)}_{12} e^{i\delta_{12}} & 0 & y^{(3)}_{23} e^{i\delta_{23}} \\
-y_{13}^{(3)} e^{i\delta_{13}} &  -y_{23}^{(3)} e^{i\delta_{23}} & 0 \\
\end{pmatrix}.
\end{equation}
On the face of it, $y^{(6)}$ therefore depends on 12 real parameters while $y^{(3)}$ depends on 6 real parameters. However, working in the SM mass basis, we can use the global baryon number symmetry of the SM Lagrangian to rotate away one phase. If we consider each diquark separately, the physical parameters are then the magnitudes of the matrix entries, and all differences of phases ($\phi_{11}-\phi_{12}$, etc). 

\subsection{SMEFT matching and operator flavour structure}
Integrating out the diquarks at tree level gives
\begin{equation}
\label{eq:Cddlagrangian}
\mathcal{L}^{d=6}\supset C_{dd}^{ijkl} \left(\bar{d}^i_R \gamma^\mu d^j_R \right) \left(\bar{d}^k_R \gamma_\mu d^l_R \right),
\end{equation}
where $i,j,k,l$ are flavour indices and 
\begin{align}
\label{eq:sextetmatching}
C_{dd}^{ijkl} & = \frac{(y^{(6)}_{ik})^*\,y^{(6)}_{jl}}{2 M_{\Phi_{(6)}}^2} \qquad \text{(sextet)},\\
C_{dd}^{ijkl} & = \frac{(y^{(3)}_{ik})^*\,y^{(3)}_{jl}}{M_{\Phi_{(3)}}^2} \qquad \text{(triplet)}. \label{eq:tripletmatching}
\end{align}
Due to the (anti)symmetry of their respective coupling matrices, at tree level only the sextet diquark can generate $\Delta F=2$ operators in which flavour changes by 2 units. These are proportional to products of its diagonal couplings:
\begin{equation}
C_{dd}^{ijij} = \frac{(y^{(6)}_{ii})^*\,y^{(6)}_{jj}}{2 M_{\Phi_{(6)}}^2} \qquad \text{(sextet)}.
\end{equation}

\subsection{Example UV completion: Goldstone bosons arising from new strong dynamics}
\label{sec:comphiggs}
Although our main focus is a general study of these diquark states, we take a detour here to discuss an example UV completion and the coupling structure it could predict for the model. Since they are scalar particles, a plausible UV origin of the diquarks is as pseudo-Goldstone bosons within a composite Higgs theory. In this scenario, both the Higgs doublet and a diquark are naturally lighter than other states in the theory since they both arise as pseudo-Goldstone bosons of a spontaneously broken global symmetry $\mathcal{G}/\mathcal{H}$ in a new strong sector~\cite{Kaplan:1983fs}. The paradigm of partial compositeness~\cite{Kaplan:1991dc}, which provides an explanation for the Higgs Yukawa couplings, then implies a flavour structure on the diquark couplings to the SM quarks.

The SM quantum numbers of the pseudo-Goldstone bosons depend on the coset of the broken symmetry group, and how the SM gauge groups are embedded within it. To build a coset that produces only the Higgs doublet and a diquark, we can start from the minimal composite Higgs model (MCHM)~\cite{Agashe:2004rs}, in which the Higgs doublet $H$ arises from the spontaneous breaking of $SO(5)$ to $SU(2)_L \times SU(2)_R$, with $H$ transforming as a bidoublet of the unbroken subgroup. We need to extend the coset space to also produce a diquark and its antiquark. For the sextet, this can be done with $Sp(6)$ broken to $SU(3)_c \times U(1)_s$.\footnote{We thank Joe Davighi for pointing us to the $Sp(6)$ group for this.} The broken generators transform as $\mathbf{\bar 6}_{-2}$ and $\mathbf{6}_{2}$ under the unbroken subgroup.
Therefore, the coset that can produce the sextet diquark and a Higgs doublet is: 
\begin{equation}
\label{eq:cosetsextet}
\frac{Sp(6) \times SO(5)}{SU(3)_c \times U(1)_s \times SU(2)_L\times SU(2)_R}.
\end{equation}
The hypercharge gauge group is embedded as $T_Y= -\frac{1}{3}T_s +T_{3R} +T_X$ where $T_s$ generates $U(1)_s$, $T_{3R}$ belongs to the $SU(2)_R$ algebra, and $T_X$ generates an additional $U(1)_X$ (under which the diquark and Higgs are uncharged), required to reproduce the correct SM hypercharge assignments. More details on the global symmetry breaking can be found in App.~\ref{sec:cosetapp}.

Instead, to generate the triplet diquark, the MCHM coset can be extended with $SO(6)$ breaking to $SU(3)_c \times U(1)_t$. The broken generators then transform as $\mathbf{\bar 3}_{-4}$ and $\mathbf{3}_{4}$ under the unbroken subgroup. So the coset that can produce the triplet diquark and a Higgs doublet is:
\begin{equation}
\label{eq:cosettriplet}
\frac{SO(6) \times SO(5)}{SU(3)_c \times U(1)_t \times SU(2)_L\times SU(2)_R}.
\end{equation}
The hypercharge gauge group in this case is embedded as $T_Y= \frac{1}{6}T_t +T_{3R} +T_X$ where $T_t$ generates $U(1)_t$, $T_{3R}$ belongs to the $SU(2)_R$ algebra, and $T_X$ generates an additional $U(1)_X$ (under which the diquark and Higgs are uncharged), required to reproduce the correct SM hypercharge assignments. See App.~\ref{sec:cosetapp} for more details on the global symmetry breaking.

The global symmetry $\mathcal{G}$ is explicitly broken by the gauging of the SM subgroup, as well as by the linear couplings between the elementary and composite sectors. Via this breaking, the pseudo-Goldstone bosons obtain a mass term. To comply with collider constraints (see next Section), the diquark is expected to have a mass rather larger than the Higgs mass, $\gtrsim 1$ TeV. Although they come from the same underlying dynamics, this hierarchy is plausibly achievable due to the QCD charge of the diquark, and the fact that the Higgs and diquark masses arise from explicit breakings of distinct parts of the coset. Studies of similar composite Higgs setups with enlarged cosets have found that PNGBs with colour charge receive additional mass contributions from QCD loops, rendering them heavier than their colour-neutral counterparts~\cite{Gripaios:2009dq,Belyaev:2016ftv}. In this way, the natural size of the diquark mass for an untuned Higgs vev can be expected to be close to the TeV scale. Furthermore, given the current precision on Higgs coupling measurements, it is now necessary in MCHM-type models that there is some tuning in the Higgs mass, meaning the natural size of the diquark mass could be relatively higher. More quantitative statements on the spectrum of PNGBs in composite Higgs models could be made in future via lattice QCD calculations~\cite{DelDebbio:2017ini,Ayyar:2018zuk,Ayyar:2018glg,DeGrand:2019vbx}, but here we simply assume the possibility of a mild mass hierarchy between the Higgs boson and the diquarks, and treat the diquark mass as a free parameter.

In the composite setup considered here, the couplings of each diquark can be set by the effective paradigm of partial compositeness~\cite{Kaplan:1991dc}. Here, the Yukawa couplings of the SM fermions arise due to linear couplings between elementary states $f^a$ (with $a\in \{Q,u,d,L,e\}$) and fermionic operators of the strong sector $\mathcal{O}^a$. After electroweak symmetry breaking, the light eigenstates which correspond to the SM fields are then linear combinations of fundamental and composite states:
\begin{align}
\label{eq:PCmixings}
f_{SM}^a&= \cos\theta^a f^a +\sin\theta^a \mathcal{O}^a,
\end{align}
such that $\sin\theta^a\sim \epsilon^a$ is the degree of compositeness of the SM fields. So projecting operators such as $g_{\rho} \bar{\mathcal{O}}^Q H\mathcal{O}^u$ (where $g_\rho$ is a strong sector coupling) along the SM fields, we can read off the strength of the Yukawa interactions.\footnote{In order to write down these couplings of the PNGBs and the SM fermions, the elementary fields must be embedded into suitable representations of the group $\mathcal{H}$. More details on this are provided in App.~\ref{sec:cosetapp}.} For the quarks, these are:
\begin{align}
(Y_u)_{ij} \sim g_\rho \epsilon^Q_i\epsilon^u_j, ~~~(Y_d)_{ij} \sim g_\rho \epsilon^Q_i\epsilon^d_j.\label{eq:yuksPC}
\end{align}
The $\epsilon^Q_i$, $\epsilon^d_i$ and $\epsilon^u_i$ can be chosen to reproduce the hierarchies of quark masses and mixings:
\begin{align}
\label{eq:PCYuks}
m_i^u \sim g_\rho \frac{v}{\sqrt{2}} \epsilon_i^Q \epsilon_i^u, ~~m_i^d \sim g_\rho \frac{v}{\sqrt{2}} \epsilon_i^Q \epsilon_i^d,~~
\frac{\epsilon_1^Q}{\epsilon_2^Q}\sim \lambda,~~\frac{\epsilon_2^Q}{\epsilon_3^Q}\sim \lambda^2,
\end{align}
where $\lambda\sim 0.23$ is the Cabibbo angle. The relations \eqref{eq:PCYuks} consist of 8 independent relations, which can be used to reduce the 10 parameters that describe the quark Yukawa sector in our framework ($g_\rho$, $\epsilon_i^Q$, $\epsilon_i^u$, $\epsilon_i^d$) down to two, which we choose to be $g_\rho$ and $\epsilon_3^Q$. Then, substituting in the measured values for the quark masses and mixings, the compositeness parameters of the down-type quarks are~\cite{Gripaios:2014tna}:
\begin{align}
\epsilon_1^d&=1.24 \times 10^{-3}/(g_\rho \epsilon_3^Q),\nonumber \\
\epsilon_2^d&=5.29 \times 10^{-3}/(g_\rho \epsilon_3^Q),\label{eq:dmixingsCH}\\\epsilon_3^d&=1.40 \times 10^{-2}/(g_\rho \epsilon_3^Q).\nonumber
\end{align}
Since the diquarks are also composite states, their coupling to down-type quarks depends on these parameters via:
\begin{equation}
\label{eq:PCcouplings}
y_{ij}^{(6)}=c^{(6)}_{ij} g_\rho \epsilon_i^d \epsilon_j^d, ~~y_{ij}^{(3)}=c^{(3)}_{ij} g_\rho \epsilon_i^d \epsilon_j^d,
\end{equation}
where $c^{(6)}_{ij}$ and $c^{(3)}_{ij}$ are unknown $O(1)$ complex parameters. These $c^{(6,3)}_{ij}$ must carry the symmetry properties of the overall diquark coupling matrices, i.e.~$c^{(6)}_{ji}=c^{(6)}_{ij}$, $c^{(3)}_{ji}=-c^{(3)}_{ij}$. Overall, the parametric dependence of the coupling matrices is $y_{ij} \propto (g_\rho (\epsilon_3^Q)^2)^{-1}$, multiplied by dimensionless numbers determined by Eqs.~\eqref{eq:dmixingsCH} and the $c_{ij}$ coefficients. Demanding that $\epsilon_3^u<1$ and $g_\rho\lesssim 4\pi$ results in a lower bound on $\epsilon_3^Q$ from the requirement to reproduce the top Yukawa \eqref{eq:yuksPC}, constraining it within the approximate range:
\begin{equation}
\frac{\sqrt{2} m_t}{4\pi v} \lesssim \epsilon_3^Q < 1.
\end{equation}
This then defines a range in the parameter combination appearing as a factor in the $y_{ij}$ couplings:
\begin{equation}
\label{eq:PCtheoryconstr}
\frac{1}{4\pi} \lesssim (g_\rho (\epsilon_3^Q)^2)^{-1} \lesssim 4\pi.
\end{equation} 
Since the diquarks only couple to down-type quarks, the partial compositeness paradigm produces a suppression in all their couplings, particularly in their flavour-changing couplings.

As we have seen, partial compositeness involves a range of baryon-like states in the composite sector (denoted by the operators $\mathcal{O}_a$ in \eqref{eq:PCmixings}) that lift the elementary fermions to their observed masses. In particular, the $O(1)$ Yukawa coupling of the top quark implies that the corresponding QCD-triplet composite states $T$ should be relatively light, and can be efficiently searched for at hadron colliders, employing similar techniques to those discussed in this work, albeit with a more complex decay phenomenology~\cite{DeSimone:2012fs,Azatov:2013hya,CMS:2018ubm,Banerjee:2024zvg}. Depending on their representations under the $SU(2)_L\times SU(2)_R$ group, some of these top partners can have exotic electric charges \cite{Contino:2008hi,Gripaios:2014pqa}, leading to striking signatures such as same-sign dileptons. Current searches for top quark partners are sensitive to masses below 2~TeV~\cite{ATLAS:2023pja} in the $T\to tZ$ channel, which rises to around 10 TeV at a future FCC-hh~\cite{Brown:2020uwk} (this reference also details modifications of the SM fermion's weak couplings). A precise computation of the mass spectrum of the composite states requires the input of lattice computations, but if the diquarks and top partners arise from the same strong underlying dynamics as contemplated in Sec.~\ref{sec:comphiggs}, their masses are broadly expected to be of the same order. Hence, a potential discovery of a top partner supplied by a positive {\emph{or}} negative outcome of searches for scalar diquarks will shed light on the nature of electroweak symmetry breaking and the flavour sector.

\section{Collider phenomenology}
\label{sec:collider}
The most direct avenue for constraining the presence of coloured objects is the direct search for resonances (``bump hunting''). Experimentally, this can be achieved in a fully data-driven approach, reducing systematics through side-band analyses. Given the scalar nature of the diquarks discussed in this work, there is a further possibility for obtaining indirect constraints from Higgs physics due to portal-type interactions. We discuss the sensitivity expected from both collider avenues in the following.

\subsection{Bump hunting}
\label{sec:bump}
Given the non-trivial colour structure of the triplet and sextet diquarks, their pair production at the LHC proceeds unsuppressed via QCD (see also Refs~\cite{Han:2009ya,Pascual-Dias:2020hxo,Bordone:2021cca,Carpenter:2021rkl,Crivellin:2022nms}). Their decays are direct probes of the flavour structure discussed earlier (assuming a large gap between diquark mass and final state)
\begin{equation}
\label{eq:partwidth}
\begin{split}
\Gamma \left(\Phi_{(3)}\to d_i d_j \right) &\simeq  {1\over 3} {3 |y_{ij}^{(3)}|^2 \, M_{\Phi_{(3)}} \over 2 \pi },\\
\Gamma \left(\Phi_{(6)} \to d_i d_j \right) &\simeq  {1\over 6} {3|y_{ij}^{(6)}|^2\, M_{\Phi_{(6)}} \over 2\pi} { 1 \over  1 + \delta_{ij} },
\end{split}
\end{equation}
where we have made the colour averaging explicit for $\Phi_{(3)},\Phi_{(6)}$ and have exploited (anti) symmetry of the Yukawa couplings (as $y^{(3)}_{ij}$ is antisymmetric $\Gamma (\Phi_{(3)}\to d_i d_j )$ vanishes for $i=j$). Through crossing symmetry, the single production of the triplet and sextet states is governed by the same matrix elements. Branching ratios for our resonance analysis follow directly from Eq.~\eqref{eq:partwidth}
\newcommand{\BR}{\text{BR}}
\begin{equation}
{\BR}(\Phi_{(i)} \to d_k d_l) = {\Gamma \left(\Phi_{(i)} \to d_k d_l \right) \over \sum_{q,m}  {\Gamma \left(\Phi_{(i)} \to d_q d_m \right)}}\,.
\end{equation}
On the hadron collider production side at leading order, when neglecting quark masses in the parton model, the single production cross sections as a result of Eq.~\eqref{eq:partwidth} scale as $\sigma (\Phi_{(3)}) = \sigma ( \Phi_{(6)} ) $ for identical couplings $y^{(3)}_{ij}=y^{(6)}_{ij} (i\neq j)$ and masses $M_{\Phi_{(3)}}=M_{\Phi_{(6)}}$ in the convention of Eq.~\eqref{eq:conventions}. The total widths differ by a factor of two $\Gamma (\Phi_{(3)}) = 2 \Gamma (\Phi_{(6)})$. Higher-order QCD corrections will lead to a departure of this tree-level result~\cite{Han:2009ya} at the 10\% level. This will not impact our findings and, hence, we will ignore these effects in the following.  It is clear, however, that any change in rate can be absorbed into the measurement of the Yukawa couplings. The integrated cross section alone is therefore not enough information to discriminate the triplet from the sextet in case a discovery is made in the singly-resonant channels alone. This holds, in particular, if the analysis is not flavour-tagged (for instance, a discovery with a double $b$-tag would directly rule out the triplet).\footnote{Of course, if an anomaly is detected, a range of flavour-tagged analyses are available to constrain the parameter space and cross-validate the anomaly in statistically orthogonal regions. We have checked that $b$-tagged analyses~\cite{CMS:2022eud} do not provide superior sensitivity compared to the flavour-agnostic bump hunt.}

\begin{figure}[!t]
\centering
\subfigure[\label{fig:lhc_dijet_constraints} LHC singly-resonant diquark production.]{\includegraphics[width=0.48\linewidth]{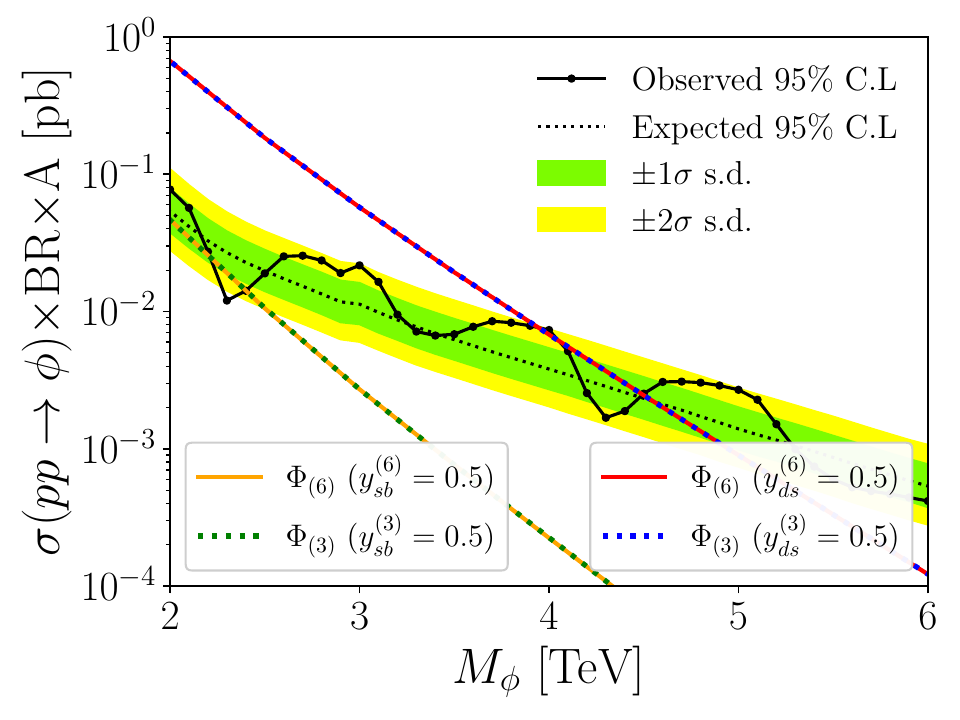}}\hfill
\subfigure[\label{fig:lhc_pair_constraints} LHC doubly-resonant diquark production.]{\includegraphics[width=0.48\linewidth]{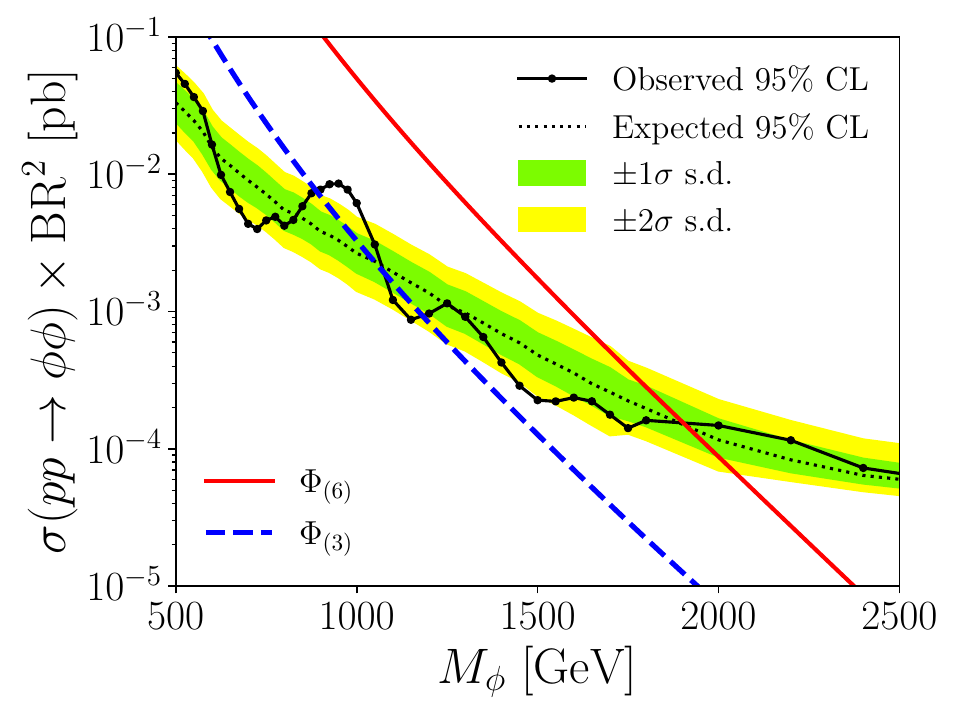}}%
\caption{Cross section of the production of (a) a single scalar diquark ($\phi=\Phi_{(6)},~\Phi_{(3)}$), and (b) pair of scalar diquarks, each decaying into two jets at the LHC ($\sqrt{s} = 13~\text{TeV}$). In each case, the diquark has only a single coupling to quarks, with all other couplings set to zero. Pair production of resonances in (b) is dominated by gluon initial states. We include the observed and expected $95\%$ CL upper limits on $\sigma \times \BR \times \text{Acceptance}$ for dijet resonances decaying into quark pairs from~\cite{CMS:2019gwf}. The observed and expected $95\%$ CL upper limits on $\sigma \times \BR^2$ for doubly-resonant production of pairs of dijet resonances~\cite{CMS:2022usq} at the LHC ($13~\text{TeV}$) is overlaid in (b).}\label{fig:lhc_constraints}%
\end{figure}

In the following, we will not include final state flavour information. To derive constraints from resonance searches, at present LHC (13 TeV), future HL-LHC (14 TeV) and a potential FCC-$hh$ (100 TeV), we generate events $ p p \rightarrow \Phi_{(i)} \rightarrow j j$ ($i=3,6$), using {\sc{FeynRules}}~\cite{Christensen:2008py,Alloul:2013bka}, {\sc{Ufo}}~\cite{Degrande:2011ua}, and \mg~\cite{Alwall:2014hca, Hirschi:2015iia}. We will give exclusion contours for representative coupling choices below, but our implementation is entirely general thus enabling a complete and direct comparison with flavour data in Sec.~\ref{sec:flavourcoll}. Using existing experimental searches as a reference point (Ref.~\cite{CMS:2019gwf}), we estimate acceptances following the prescription of \cite{Pascual-Dias:2020hxo,CMS:2018mgb}, resulting in an acceptance of 0.57 for diquark masses above 1.6 TeV. 

\begin{figure}[!t]
\centering
\subfigure[\label{fig:hllhc_constraints}HL-LHC 14 TeV collisions.]{\includegraphics[width=0.48\textwidth]{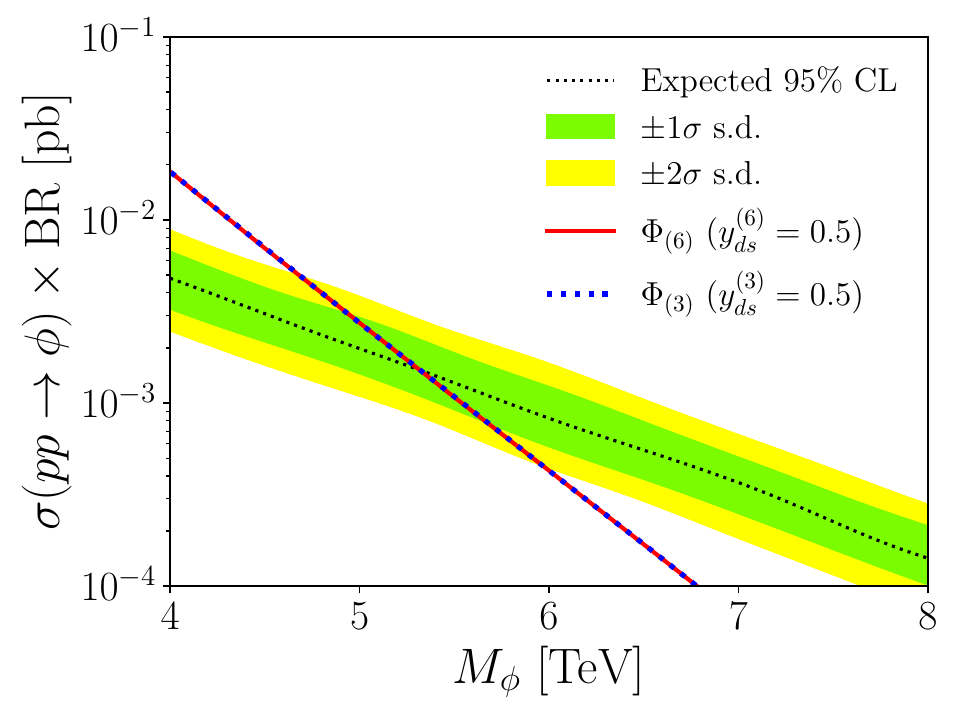}}\hfill
\subfigure[\label{fig:fcc_constraints}FCC-$hh$ 100 TeV collisions.]{\includegraphics[width=0.48\textwidth]{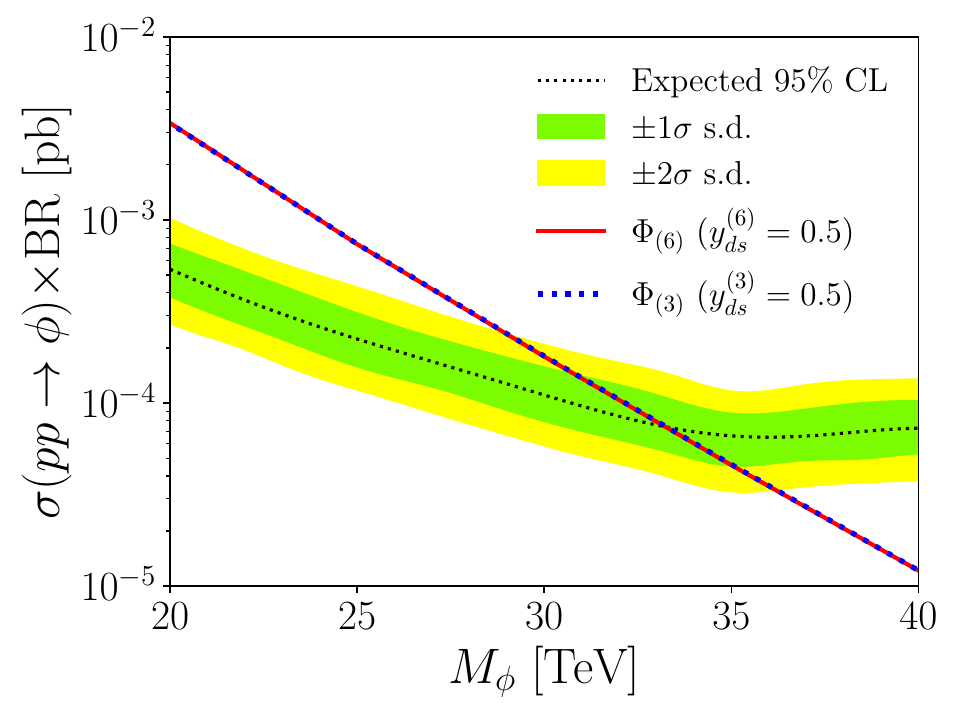}}
\caption{Cross section of the production of a single scalar diquark ($\phi=\Phi_{(6)},~\Phi_{(3)}$) decaying into two jets at (a) $\sqrt{s} = 14~\text{TeV}$ (HL-LHC), and (b) $\sqrt{s} = 100~\text{TeV}$ (FCC-$hh$). The expected $95\%$ CL upper limits on $\sigma \times \BR$ for dijet resonances at the HL-LHC~\cite{Chekanov:2017pnx} and the FCC-$hh$~\cite{Helsens:2019bfw} are overlaid in the respective plots. A vastly increased centre-of-mass energy enables a much wider exploration of the coloured resonance parameter range.}\label{fig:lhc_constraints2}
\end{figure}

For LHC and HL-LHC collisions, events are generated for resonance masses in the range $2-8~\text{TeV}$, and the resulting cross section is plotted as a function of $M_\phi$ (with $\phi=\Phi_{(6)},~\Phi_{(3)}$) in Fig.~\ref{fig:lhc_dijet_constraints} for the LHC and in Fig.~\ref{fig:hllhc_constraints} for the HL-LHC. The observed and expected $95\%$ confidence levels (C.L.) on dijet resonance searches at the LHC~\cite{CMS:2019gwf} are included from \textsc{HepData}~\cite{Maguire:2017ypu}, which are then overlaid in Fig.~\ref{fig:lhc_dijet_constraints}. Similarly, the expected $95\%$ C.L.~constraints at the HL-LHC~\cite{Chekanov:2017pnx} are overlaid in Fig~\ref{fig:hllhc_constraints}. For the FCC-$hh$ extrapolation, we scan over masses from $20-40~\text{TeV}$ and plot them in Fig.~\ref{fig:fcc_constraints}. The projected FCC-$hh$ $95\%$ C.L.~constraints from Ref.~\cite{Helsens:2019bfw} are overlaid in Fig.~\ref{fig:fcc_constraints}. As can be seen, the LHC is capable of exploring mass scales of the order 7~TeV, depending on the coupling choices. This is vastly improved at a purpose-built FCC-hh that can probe resonances of $\sim 40$~TeV, again depending on the specific coupling choices (Fig.~\ref{fig:lhc_constraints2}).

For searches for the double-resonant production of pairs of dijet resonances (see also~\cite{Schumann:2011ji}), we generate events for the process $p p \rightarrow \phi \phi \rightarrow j j j j$ ($\phi = \Phi_{(6)},~\Phi_{(3)}$) at $\sqrt{s}=13~\text{TeV}$, and scan over masses from $0.5-3~\text{TeV}$, and plot the corresponding cross sections in Fig.~\ref{fig:lhc_pair_constraints}. The current observed and expected $95\%$ C.L. constraints at the LHC~\cite{CMS:2022usq} are again imported from \textsc{HepData} and overlaid in this figure. From Fig.~\ref{fig:lhc_dijet_constraints}, we see that the constraining power of the singly resonant dijet searches varies depending on the couplings of the diquarks, loosening significantly if the diquark's only couplings involve a $b$ quark, due to the low $b$ parton luminosity. Since resonance pair production proceeds through pure QCD interactions, it cannot be weakened by suppressing couplings and can be considered a baseline constraint. The diquark decay lengths are generally very small for the coupling sizes we consider here; even for couplings suppressed by partial compositeness as in Sec.~\ref{sec:comphiggs}, their decays are prompt. On the other hand, if the resonances are ultra-weakly coupled, displaced jet searches could become sensitive for a very narrow range of the couplings that equates to an average lifetime that matches the vertex detector location of the LHC multi-purpose experiments. We will not consider this latter avenue any further in this work.

\subsection{Higgs signal strengths}
As charged scalars, the sextet and triplet states enable renormalisable portal interactions with the Higgs sector via interactions 
\begin{equation}
{\cal{L}}\supset -\sum_{i=3,6} \lambda_{Hi} \left(\Phi_H^\dagger \Phi_H\right) \left(\Phi_{(i)}^\dagger \Phi_{(i)}\right)
\end{equation}
(where $\Phi_H$ is the SM Higgs doublet) as well as quartic interactions of the diquarks among themselves that we are not considering in this work. After electroweak symmetry breaking, these quartic interactions lead to new contributions to Higgs production, e.g. to the dominant $gg\to H$ production mode and to loop-induced Higgs decays $H\to gg, \gamma\gamma, \gamma Z$. These contributions give rise to indirect constraints on the model space\footnote{Another important set of indirect constraints for BSM physics are electroweak precision data. As already mentioned in Sec.~\ref{sec:down}, $\Phi_{(3),(6)}$ are $SU(2)_L$ singlets, and they do not contribute significantly to oblique corrections. Electroweak precision data are largely insensitive to the mass scales that we focus on in this work.} and follow the Appelquist-Carazzone decoupling pattern~\cite{Appelquist:1974tg}. The current Higgs signal strength observation by ATLAS and CMS can be reproduced already for moderately light diquark masses.

\begin{figure}[!t]
\centering
\subfigure[\label{fig:higgs_brs}]{\includegraphics[width=0.48\textwidth]{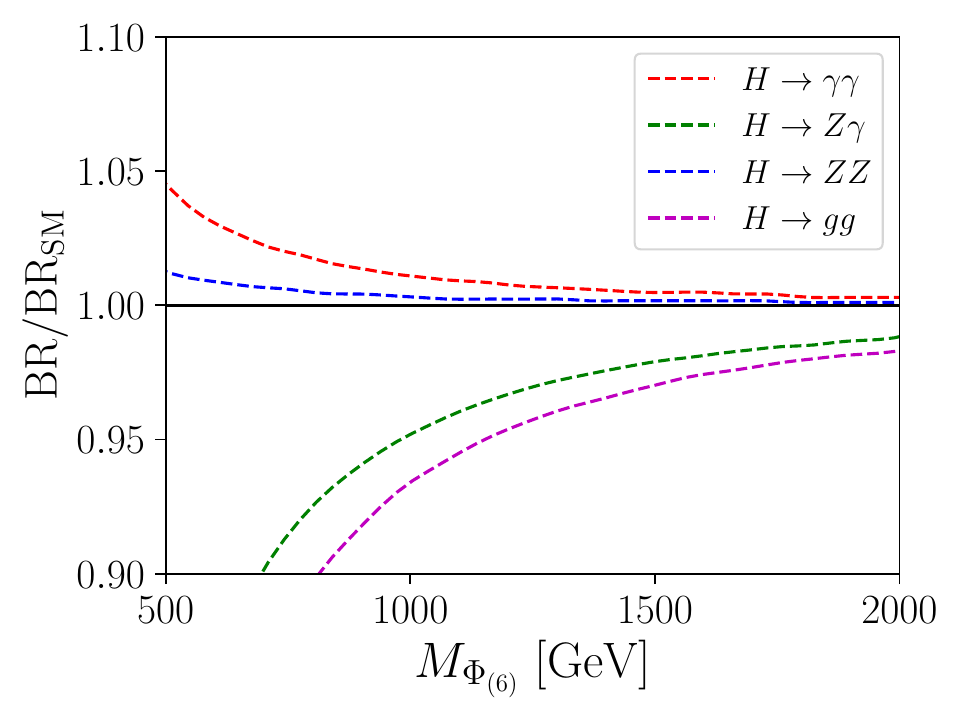}}
\subfigure[\label{fig:higgs_constraints}]{\includegraphics[width=0.48\textwidth]{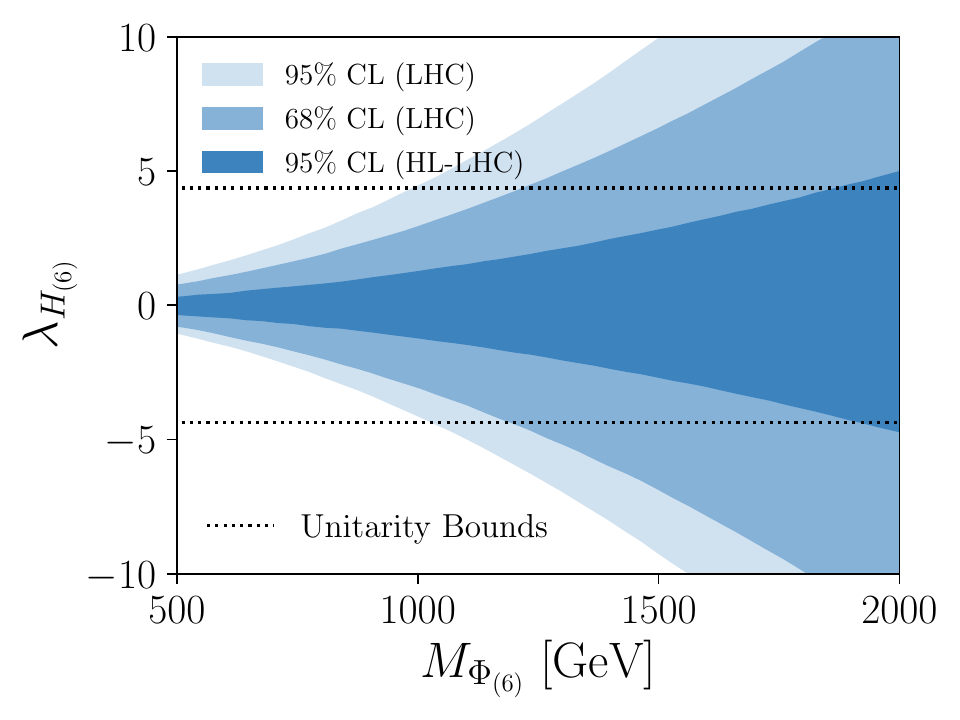}}\hfill
\caption{(a) Modified Higgs branching ratios as a function of $M_{\Phi_{(6)}}$ for $\lambda_{H_{(6)}}=1$.  (b) $95 \%$ and $68 \%$ confidence level constraints on $\lambda_{H_{(6)}}$ and $M_{\Phi_{(6)}}$ from Higgs signal strengths~\cite{ATLAS:2021vrm,ATLAS:2022vkf} at the LHC. The extrapolated $95\%$ constraints at the HL-LHC~\cite{Cepeda:2019klc}, as well as the constraints from unitarity, are overlaid. For masses $M_{\Phi_{(6)}}\gtrsim 2~\text{TeV}$ that the LHC is exploring, Figs~\ref{fig:lhc_constraints} and \ref{fig:lhc_constraints2}, Higgs constraints do not provide additional sensitivity.\label{fig:Higgs}}
\end{figure}

In Fig.~\ref{fig:higgs_brs} we show the corrections to Higgs branching ratios due to the sextet diquark, taking the SM decay widths from~\cite{LHCHiggsCrossSectionWorkingGroup:2011wcg}. In Fig.~\ref{fig:higgs_constraints}, we show the resulting constraints from the Higgs measurements of Refs.~\cite{ATLAS:2021vrm,ATLAS:2022vkf}, where the leading constraints are coming from $H\to \gamma\gamma,ZZ$. We also include constraints from unitarity considerations (employing a partial wave analysis of $H\Phi_{(6)}$ forward scattering) of the extended scalar sector and a signal strength projection for the HL-LHC (using the results of~\cite{deBlas:2019rxi} which set signal strength modifications to $\lesssim 5\%$ depending on the Higgs production and decay channel). For the mass scales explored at LHC, we conclude that Higgs physics does not increase the new physics discovery potential. 

\section{Flavour phenomenology}
\label{sec:flavour}
In the previous section, we saw that the sextet and triplet diquark have rather similar collider phenomenology, as long as they have couplings to light quarks. In contrast, their leading flavour phenomenology can be drastically different, due to the different symmetry properties of their coupling matrices, Eq.~\eqref{eq:couplingmatrices}. 

\subsection{Diagonal couplings of the sextet diquark}
\label{sec:diagflavour}
\begin{figure}[!t]
\begin{center}
\begin{tikzpicture}
   \begin{feynman}
       \vertex (d1) at (-2, 0) {\(\large d_i\)};
       \vertex (d2) at (2, 0) {\(\large \bar{d}_i\)};
       \vertex (a) at (0, 0);
       \vertex (a1) at (0,0.4) {$\large y_{ii}^{(6)}$};
       \vertex (b) at (0., -2.);
       \vertex (b1) at (0,-2.4) {$\large y_{jj}^{(6) *}$};
       \vertex (d3) at (-2, -2) {\(\large \bar{d}_j\)};
       \vertex (d4) at (2, -2) {\(\large d_j \)};
       
       \diagram* {
           (d1) -- [fermion] (a) -- [anti fermion,edge label=] (d2),
           (a) -- [scalar, edge label=\(\Phi_{(6)}\)] (b),
           (b) -- [fermion] (d3),
           (b) -- [fermion, edge label=] (d4),
       };
   \end{feynman}
   \fill[black] (0, 0) circle (2pt);
   \fill[black] (0, -2) circle (2pt);
   \begin{feynman}
       \vertex (d1) at (3.5, -0.) {\(\large d_i\)};
       \vertex (d2) at (8, -0.) {\(\large d_j\)};
       \vertex (a) at (5, -0.);
       \vertex (a1) at (5,0.3) {$\large y_{ik}$};
       \vertex (c) at (6.5, -0.);
       \vertex (c1) at (6.5,0.35) {$\large y^{*}_{jk}$};
       \vertex (b) at (5,-2);
       \vertex (a1) at (5,-2.35) {$\large y^{*}_{jk}$};
       \vertex (d) at (6.5,-2);
       \vertex (d0) at (6.5,-2.3) {$\large y_{ik}$};
       \vertex (d3) at (3.5, -2) {\(\large \bar{d}_j\)};
       \vertex (d4) at (8, -2) {\(\large \bar{d}_i \)};
       
       \diagram* {
           (d1) -- [fermion] (a) -- [anti fermion, edge label=$d_k$] (c) -- [fermion] (d2),
           (a) -- [scalar, edge label'=\(\Phi\)] (b),
           (c) -- [scalar, edge label=\(\Phi\)] (d),
           (d3) -- [anti fermion] (b) -- [fermion, edge label' = $d_k$] (d) -- [anti fermion] (d4),
       };
   \end{feynman}
   \fill[black] (5, 0) circle (2pt);
   \fill[black] (5, -2) circle (2pt);
   \fill[black] (6.5, 0) circle (2pt);
   \fill[black] (6.5, -2) circle (2pt);
\end{tikzpicture}
\caption{Diquark diagrams contributing to neutral meson mixing. Left: tree diagram involving only diagonal couplings. Right: loop diagram involving only off-diagonal couplings ($i\neq j \neq k$). \label{fig:mixing}
}
\end{center}
\end{figure}
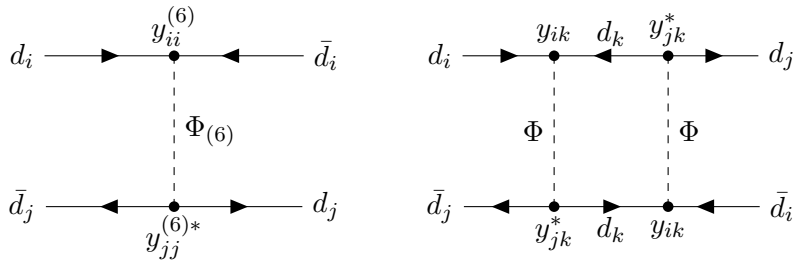

The mixing of neutral $B$ mesons provides an important test of new physics in down-type four-quark operators. The sextet diquark can contribute to neutral down-type meson mixing via the left-hand diagram in Fig.~\ref{fig:mixing}, which depends only on its diagonal couplings. Loop-level effects proportional to off-diagonal couplings also arise for both diquarks (right diagram in Fig.~\ref{fig:mixing}), which we will make use of later. 

The mass differences of the neutral B mesons are measured precisely~\cite{ParticleDataGroup:2024cfk}
\begin{align}
\Delta M_s=17.765 \pm 0.006~\text{ps}^{-1},~~~\Delta M_d=0.533^{+0.022}_{-0.036}~\text{ps}^{-1}.
\end{align}
For the Standard Model prediction we take the weighted average of Ref.~\cite{DiLuzio:2019jyq}, which averages the results of~\cite{FlavourLatticeAveragingGroup:2019iem,Grozin:2018wtg,Kirk:2017juj,King:2019lal,Dowdall:2019bea,Boyle:2018knm}:
\begin{align}
\Delta M_s^{\text{SM}}=18.4^{+0.7}_{-1.2}~\text{ps}^{-1},~~~\Delta M^{\text{SM}}_d=(0.5065 \pm 0.0019)~\text{ps}^{-1}.
\end{align}
New physics can enter into $B_q$ mixing observables (with $q=s,d$) through the quantities $\Delta_q$ and $\phi^{\Delta}_q$, defined by~\cite{Lenz:2006hd}:
\begin{equation}
    M_{12}^q = M_{12}^{\text{SM},q} \, \Delta_q, \hspace{3cm} \Delta_q = |\Delta_q| e^{i \phi_q^{\Delta}}.
\end{equation}
In terms of the Wilson coefficient $C^{ijkl}_{dd}$ these are~\cite{DiLuzio:2019jyq}:
\begin{equation}
\Delta_q=1-\eta^{6/23} \frac{\sqrt{2}C^{qbqb}_{dd}}{4 G_F (V_{tb} V_{tq}^*)^2R^{\text{SM}}_{\text{loop}}}.
\end{equation}
where $\eta=\alpha_s(\mu_{NP})/\alpha_s(m_b)$ and $R^{\text{SM}}_{\text{loop}}=(1.310 \pm 0.010)\times 10^{-3}$ is the Standard Model loop function~\cite{DiLuzio:2019jyq}. 
The mass difference, width difference, and semileptonic CP asymmetry can then be written in terms of $\Delta_q$ and $\phi_q^{\Delta}$ as~\cite{Lenz:2006hd}
\begin{align}
 \Delta M_q &= \Delta M_q^{\text{SM}} |\Delta_q|, \label{eq:deltaMsNP}\\
     \Delta \Gamma_q &= 2 |\Gamma_{12}^q| \cos(\phi_q^{\text{SM}} + \phi_q^{\Delta}), \label{eq:delta_gamma} \\
    a_{\text{fs}}^q &= \frac{|\Gamma_{12}^q|}{|M^{\text{SM},q}_{12}|} \frac{\sin(\phi^{\text{SM}}_q + \phi^{\Delta}_q )}{|\Delta_q|}, \label{eq:a_fs}
\end{align}
where we have assumed that $\Gamma_{12}$ does not change from its SM value, since it is dominated by decays to which the diquarks do not contribute at tree level.
In general, the three observables $\Delta_q$, $\Gamma_q$ and $a_{\text{fs}}^q$ are independent, however, given the lack of contribution to $\Gamma^q_{12}$, effectively only two out of the three are independent within the diquark parameter space. 

\begin{figure}[!t]
\centering
\subfigure[\label{fig:osci_Bs}$B_s$ oscillation constraints.]{\includegraphics[width=0.48\textwidth]{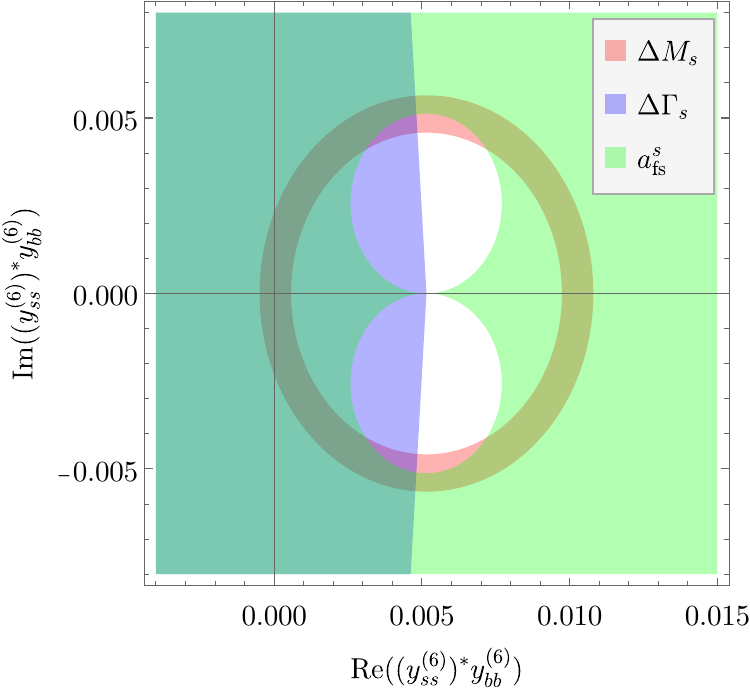}}\hfill
\subfigure[\label{fig:osci_Bd}$B_d$ oscillation constraints.]{\includegraphics[width=0.48\textwidth]{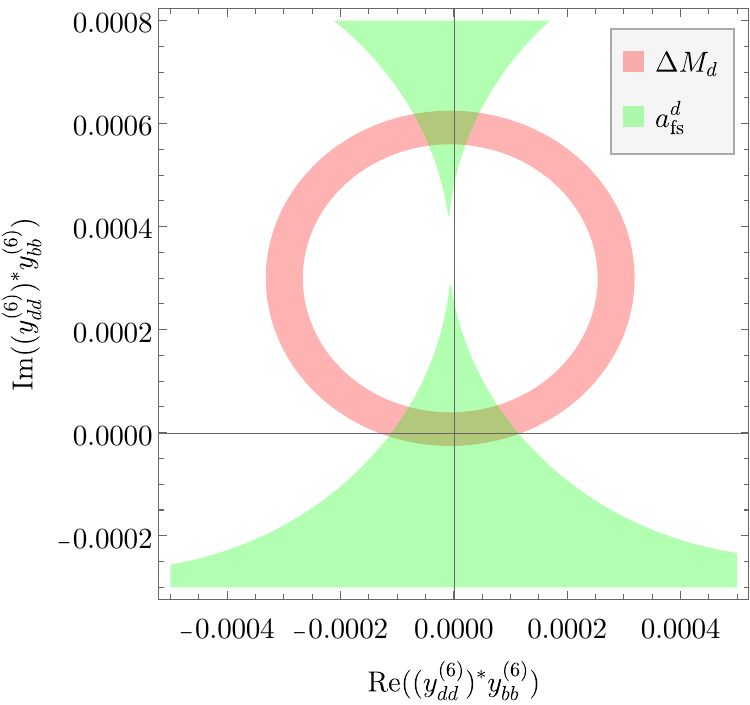}} \\ 
\caption{\label{fig:oscillations}Allowed regions (2$\sigma$) in diagonal coupling strength products from $B_s$ and $B_d$ meson oscillations for a sextet diquark mass of $M_{\Phi_{(6)}}=$5 TeV.}
\end{figure}

The experimental values for the width difference and the semileptonic asymmetries are~\cite{HFLAV:2022pwe,ParticleDataGroup:2024cfk}:
\begin{align}
a^d_{\text{fs}}&=-0.0021 \pm 0.0017,\\
a_{\text{fs}}^s&=-0.0006 \pm 0.0028, & \Delta \Gamma_s = (0.084 \pm 0.005)~\text{ps}^{-1},
\end{align}
where we have not included the $B_d$ width difference since it is not currently well measured. The SM predictions are~\cite{Lenz:2006hd}
\begin{align}
a^{d, \text{SM}}_{\text{fs}}&=(-4.8^{+1.0}_{-1.2}) \times 10^{-4},\\
a_{\text{fs}}^{s,\text{SM}}&=(2.06\pm 0.57) \times 10^{-5}, & \Delta \Gamma_s^{\text{SM}} = (0.096 \pm 0.039)~\text{ps}^{-1}.
\end{align}
Putting all of this together, the allowed region from each of these observables is shown in Fig.~\ref{fig:osci_Bs}  for $B_s$ mixing, for a diquark mass of 5 TeV. It can be seen that each observable depends on a different function of the real and imaginary parts of the coupling product, so the overall constraint on its absolute value in general depends on its phase. In Fig.~\ref{fig:osci_Bd}, the equivalent constraints from the observables $\Delta M_d$ and $a_{\text{fs}}^d$ measured in $B_d$ mixing are shown, again for a mass of 5 TeV. 

Neutral Kaons are the last system of oscillating mesons that we can use to find constraints on the diquark couplings. In this case, we use direct constraints on the corresponding Wilson coefficient $C_K$ as found by the {\sc{UTfit}} collaboration~\cite{utfit_collaboration_model-independent_2008}, where we can identify:
\begin{equation}
    C_K = C_{dd}^{dsds} =\frac{y_{dd}^{*(6)} \,y_{ss}^{(6)}}{2 M_{\Phi_{(6)}}^2}.
\end{equation}
The {\sc{UTfit}} collaboration lists the allowed regions at $2\sigma$ of the real part of $C_K$ to be $[-9.6,9.6] \times 10^{-13}$ GeV$^{-2}$ and the imaginary part to be $[-4.4,2.8] \times 10^{-15}$ GeV$^{-2}$~\cite{utfit_collaboration_model-independent_2008}. For a diquark mass of $5$ TeV, this gives the constraints (at $2\sigma$):
\begin{align}
\mathrm{Re}(y_{dd}^{*(6)} \,y_{ss}^{(6)})< [-2.4,2.4]\times 10^{-5}, ~~\mathrm{Im}(y_{dd}^{*(6)} \,y_{ss}^{(6)})< [-1.1,0.7]\times 10^{-7}.
\end{align}

A comment is in order regarding a feature of the plots in Fig.~\ref{fig:oscillations}. When comparing the allowed regions of both $B$ meson systems, they appear to be rotated by 90$^\circ$. This rotation is due to the dependence of $\Delta_s$ and $\Delta_d$ on different CKM matrix elements. $\Delta_s$ depends on $(V_{tb}V^*_{ts})^2$ which happens to be real in the Wolfenstein parameterisation while $\Delta_d$ depends on $(V_{tb}V^*_{td})^2$ which happens to be almost completely imaginary. Of course, a phase in either of these quantities can be rotated away by a constant baryon number rephasing of all quark fields, and they are therefore not in themselves physical. However, such a rephasing will change not only the phases in the CKM matrix but also the phases in the diquark coupling matrix, so in the diquark theory, this relative rotation of the constraint plots is physical, although the orientation of each individual plot is determined by the CKM parameterisation used. It is also worth noting that the three diquark phases corresponding to the $B_d$, $B_s$ and $K$ systems are not independent, since they depend only on two differences of phases: $\text{arg} ( y_{ii} y_{jj}^*) =\phi_{ii} - \phi_{jj}$. This means it is not possible to consistently redefine them such that they are each aligned with the phase of the relevant CKM product.

\begin{figure}[!t]
\centering
\includegraphics[width=0.5\textwidth]{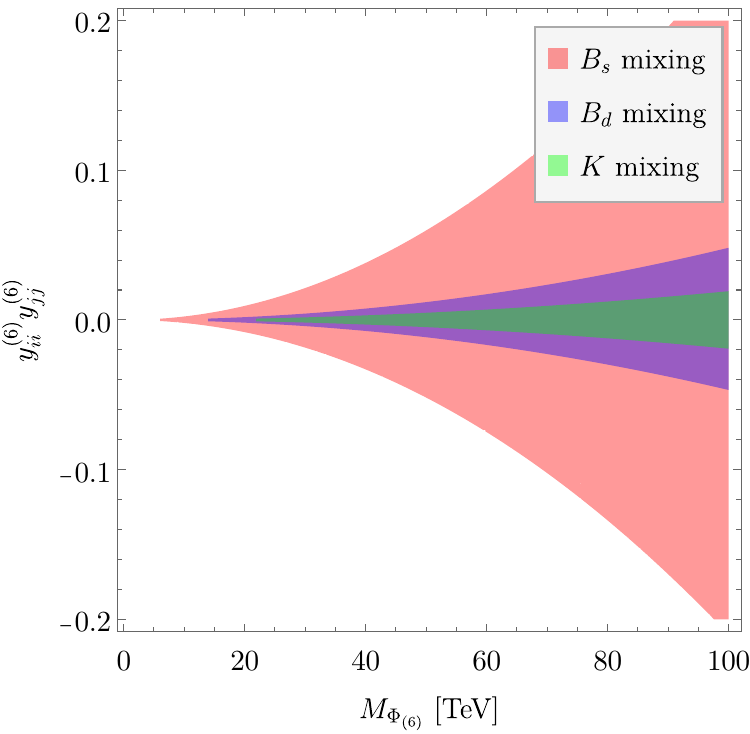}
\caption{$2\sigma$ allowed regions in diagonal coupling strength products of the sextet diquark from neutral meson oscillations, assuming real coupling strengths. The flavour indices $i$ and $j$ depend on the meson being used to set constraints, e.g.~for $K$ mixing $i,j=d,s$.}
\label{fig:oscillationsreal}
\end{figure}

\begin{figure}[!b]
\centering
\subfigure[\label{fig:CH1}Assuming real couplings]{\includegraphics[width=0.45\textwidth]{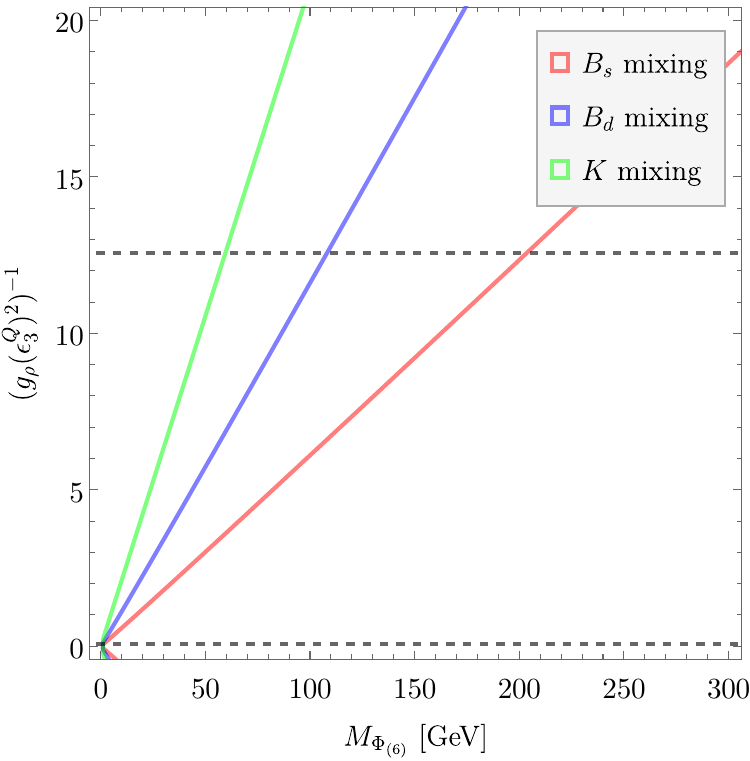}}\hfill
\subfigure[\label{fig:CH2}Most constraining complex phases]{\includegraphics[width=0.45\textwidth]{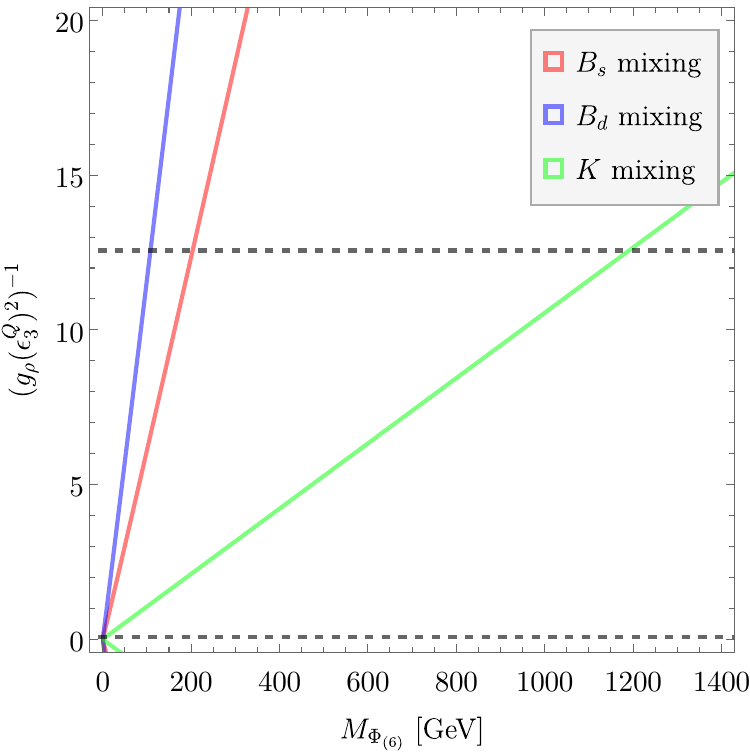}}\hfill
\caption{Constraints from meson mixing observables on the parameter space of a partially composite sextet diquark. Regions below the coloured lines are allowed (at 2$\sigma$) by current measurements. On the left plot (a), all $O(1)$ $c_{ij}$ coefficients are real and equal to one, while in the right plot (b) their phases have been varied to maximise the bounds. The region between the dotted horizontal lines is the theoretically allowed region of $(g_\rho (\epsilon_3^Q)^2)^{-1}$ (see Eq.~\eqref{eq:PCtheoryconstr}).}
\label{fig:CompHiggs}
\end{figure}

The constraints on coupling combinations from all three neutral mesons under the assumption that all couplings are real are shown in Fig.~\ref{fig:oscillationsreal}. In this case the strongest constraints from $B_s$ and $B_d$ meson mixing are from $\Delta M_s$ (where an additional wrong-sign solution is ruled out by further including $\Delta \Gamma_s$ as seen in Fig.~\ref{fig:osci_Bs}) and $a^d_{\text{fs}}$ respectively. Kaon mixing in particular is seen to impose extremely strong constraints from Re($C_K$). Looking at these in comparison to some of the current and future limits seen in Sec.~\ref{sec:collider}, searching for these diquarks at collider experiments seems like it could be a losing game.

However, we can also investigate how meson mixing constraints look within the flavour paradigm of partial compositeness, which provides a natural suppression in the diquark couplings, as described in Sec.~\ref{sec:comphiggs}. Setting all of the $O(1)$ couplings $c_{ij}^{(6)}$ to be real and equal to one, we find the allowed regions from neutral meson mixing in Fig.~\ref{fig:CH1}, bringing the allowed mass ranges well within collider sensitivity. The horizontal dashed lines delineate the theoretically allowed region of $(g_\rho (\epsilon_3^Q)^2)^{-1}$, Eq.~\eqref{eq:PCtheoryconstr}. We can also allow for complex phases in the coupling strength matrix, with the observables in meson oscillations only depending on the difference in phases between the diagonal elements again. With this, we can look at the phase configuration that leads to the strongest constraints, roughly $\phi_{bb} - \phi_{ss} = \pi$, $\phi_{bb} - \phi_{dd} = 3 \pi/2$ and $\phi_{dd} - \phi_{ss} = \pi/2$, and with $|c_{ij}^{(6)}|=1$. The resulting constraints are shown in Fig.~\ref{fig:CH2}. This time the strongest constraint on the mass $M_{\Phi_{(6)}}$ comes from Kaon oscillations, but even in this case masses of $\sim$ 1 TeV are still allowed by mixing observables across the full coupling parameter range. This exercise demonstrates that, under motivated assumptions on the couplings, meson mixing bounds can be weaker than existing collider bounds on the sextet diquark. We will explore the complementarity of collider and flavour observables further in Sec.~\ref{sec:flavourcoll}.

\subsection{Off-diagonal couplings of the sextet and triplet diquarks}
\label{sec:offdiagflavour}
\subsubsection*{Charmless $B$ decays}
At tree level, combinations of the off-diagonal couplings of both the sextet and the triplet contribute to charmless hadronic $B$ decays. 
Both diquarks match at to operators relevant for $b\to d \bar{s} s$ transitions at leading order:
\begin{align}
\label{eq:Lbtodss}
\mathcal{L}^{b\to dss}= \begin{cases}
\begin{aligned}
    \frac{(y_{sb}^{(6)})^* \,y_{sd}^{(6)}}{2 M_{\Phi_{(6)}}^2} \left(\left(\bar{b}^a_R\gamma^\mu d^a_R\right) \left(\bar{s}^b_R\gamma_\mu s^b_R\right)+\left(\bar{b}^a_R\gamma^\mu d^b_R\right)\left(\bar{s}^b_R\gamma_\mu s^a_R\right)\right), &~\quad \text{(sextet)} \\
 \frac{(y_{sb}^{(3)})^* \,y_{sd}^{(3)}}{M_{\Phi_{(3)}}^2} \left(\left(\bar{b}^a_R\gamma^\mu d^a_R\right) \left(\bar{s}^b_R\gamma_\mu s^b_R\right)-\left(\bar{b}^a_R\gamma^\mu d^b_R\right)\left(\bar{s}^b_R\gamma_\mu s^a_R\right)\right), &~\quad \text{(triplet)}
\end{aligned}
\end{cases}
\end{align}
where $a,b$ are colour indices. These operators mediate $B\to \phi\pi$, $B\to \phi\phi$ and $B\to \phi \gamma$ decays, which are extremely rare in the SM, predicted as $\BR_{\text{SM}}(B^+\to \phi\pi^+)=2.0^{+0.3}_{-0.2}\times 10^{-8}$~\cite{Bar-Shalom:2002icz}, $\BR_{\text{SM}}(B^0\to \phi\phi)=2.1^{+1.6}_{-0.3}\times 10^{-9}$~\cite{Bar-Shalom:2002icz} and $\BR_{\text{SM}}(B^0\to \phi\gamma)=3.6\times 10^{-12}$~\cite{Li:2003kz}. In the SM, the operators involved are purely left-handed, so there is no interference between these and the right-handed diquark contributions, and the diquarks can hence only enhance the branching ratios of these decays. The latest measurements of or limits on these decays are:  $\BR(B^+\to \phi\pi^+)=(3.2 \pm 1.5)\times 10^{-8}$~\cite{LHCb:2019xmb}, $\BR(B^0\to \phi\phi)<2.7\times 10^{-8}\, (90\%)$~\cite{LHCb:2019jgw} and $\BR(B^0\to \phi\gamma)<1.0\times 10^{-7}\, (90\%)$~\cite{Belle:2016tvc}. 

We use the calculations of Ref.~\cite{Bar-Shalom:2002icz} to set limits on the couplings and mass of the diquark from the most constraining decay mode, $\BR(B^+\to \phi\pi^+)$, finding:
\begin{align}
\frac{|y^{(3)*}_{sb}\,y^{(3)}_{ds}|}{M_{\Phi_{(3)}}^2} &< 4.6 \times 10^{-3}\, \text{TeV}^{-2} ~~ (95\%~\text{C.L.})\\
\frac{|y^{(6)*}_{sb}\,y^{(6)}_{ds}|}{M_{\Phi_{(6)}}^2} &< 5.1 \times 10^{-3}\, \text{TeV}^{-2} ~~ (95\%~\text{C.L.})
\end{align}

In principle, the branching ratio of the decay $B_s\to \pi\pi$, measured a few years ago by LHCb~\cite{LHCb:2016inp}, could similarly test the coupling combination $|(y_{db})^* y_{ds}|$. However the SM prediction for this decay differs significantly between the QCD factorisation and pQCD approaches~\cite{Cheng:2009mu,Xiao:2011tx,Chang:2014yma}, and we are not aware of any BSM studies of this decay, so we cannot make use of it here.

\subsubsection*{Neutral meson mixing at loop level}
Both diquarks can contribute to neutral meson mixing at loop level through their off-diagonal couplings, through the diagram shown on the right in  Fig.~\ref{fig:mixing}, and the corresponding crossed diagram. 
The loop level matching to the appropriate $\Delta F=2$ Wilson coefficients in Eq.~\eqref{eq:Cddlagrangian} gives (neglecting corrections of $O(m_{d_k}^2/M_{\Phi}^2)$):
\begin{align}
C_{dd}^{ijij}&=\frac{1}{4\pi^2M_{\Phi_{(3)}}^2}\left(y^{(3)*}_{ki} y^{(3)}_{kj} \right)^2 ~~~ \text{(triplet)},\\
C_{dd}^{ijij}&=\frac{3}{16\pi^2 M_{\Phi_{(6)}}^2}\left(y^{(6)*}_{ki} y^{(6)}_{kj} \right)^2 ~~~ \text{(sextet)},
\end{align}
where $k$ is summed over for the sextet, and necessarily $k\neq i \neq j$ for the triplet, due to its antisymmetric coupling structure.

\subsubsection*{(Semi)leptonic decays via RGEs}
Both diquarks can generate the process $d_i\to d_j \,l^+ l^-$ (and $d_i\to d_j \,\bar{\nu} \nu$) via the divergent loop diagram shown in Fig.~\ref{fig:RGEZ}. In the SMEFT language, this arises due to the operator coefficient $C_{Hd}$ being generated from $C_{dd}$ via renormalisation group running~\cite{Alonso:2013hga}, proportional to the hypercharge coupling $g_1^2$. Due to the $Z$ boson's dominant axial vector coupling to charged leptons, below the electroweak scale, the charged lepton diagram matches predominantly to the operator $O_{10}^\prime \sim \left(\bar d_i\gamma^\mu P_R d_j\right)\left(\bar l\gamma^\mu\gamma^5 l\right)$. This operator can mediate the decay $B_{s,d}\to l^+l^-$.

The blue regions in the summary plots Fig.~\ref{fig:collidervsflavour3} and Fig.~\ref{fig:collidervsflavour6} include constraints arising from these RG-induced contributions to leptonic and semileptonic $B_s$, $B_d$ and $K$ decays via the \texttt{smelli} quark likelihood~\cite{Aebischer:2018iyb,Aebischer:2018bkb,Straub:2018kue}. 

There will also be finite matching contributions, including from a diagram in which the $Z$ is radiated from the diquark. These are not included in our calculation. However, they can be expected to be subdominant to the log-enhanced divergent contributions, which as we see from the summary plots do not produce a leading constraint in any region of parameter space.

\begin{figure}
\begin{center}
\begin{tikzpicture}
    \begin{feynman}
        \vertex (d1) at (-2.25, 1) {\(\large d_i\)};
        \vertex (d2) at (-2.25, -1) {\(\large d_j\)};
        \vertex (y1) at (-1, 1);
        \vertex (y1t) at (-1,1.3) {$\large y_{ik}$};
        \vertex (y2) at (-1,-1);
        \vertex (y1t) at (-1,-1.35) {$\large y_{jk}^{*}$};
        \vertex (a) at (0, 0);
        \vertex (b) at (1.5, 0);
        \vertex (l1) at (2.5, 1) {\(\large l^{-},\nu\)};
        \vertex (l2) at (2.5, -1) {\(\large l^{+},\bar{\nu} \)};
        
        \diagram* {
            (y1) -- [anti fermion, edge label = $d_k$] (a) -- [anti fermion, edge label = $d_k$] (y2),
            (a) -- [boson, edge label=\( Z \)] (b),
            (l1) -- [anti fermion] (b) -- [anti fermion] (l2),
            (y1) -- [anti fermion] (d1),
            (y2) -- [fermion] (d2),
            (y1) -- [scalar, edge label'=$\Phi$] (y2),
        };
    \end{feynman}
    \fill[black] (0, 0) circle (2pt);
    \fill[black] (1.5, 0) circle (2pt);
    \fill[black] (-1, 1) circle (2pt);
    \fill[black] (-1, -1) circle (2pt);
\end{tikzpicture}
\caption{\label{fig:RGEZ} Divergent diagram for the contribution of either diquark to $d_i\to d_j \,l^+ l^-$ or $d_i\to d_j \,\bar\nu \nu$.}
\end{center}
\end{figure}
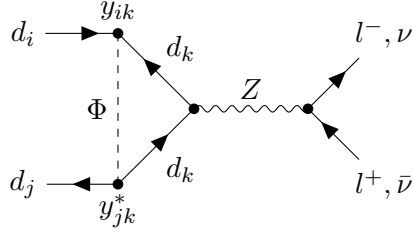
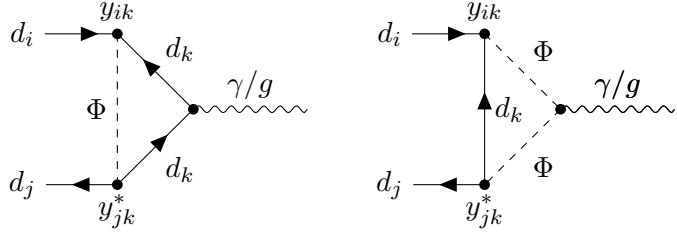
\begin{figure}
\begin{center}
\begin{tikzpicture}
    \begin{feynman}
        \vertex (d1) at (-2.25, 1) {\(\large d_i\)};
        \vertex (d2) at (-2.25, -1) {\(\large d_j\)};
        \vertex (y1) at (-1, 1);
        \vertex (y1t) at (-1,1.3) {$\large y_{ik}$};
        \vertex (y2) at (-1,-1);
        \vertex (y1t) at (-1,-1.35) {$\large y_{jk}^{*}$};
        \vertex (a) at (0, 0);
        \vertex (b) at (1.5, 0);;
        
        \diagram* {
            (y1) -- [anti fermion, edge label = $d_k$] (a) -- [anti fermion, edge label = $d_k$] (y2),
            (a) -- [boson, edge label=\( \gamma / g \)] (b),
                        (y1) -- [anti fermion] (d1),
            (y2) -- [fermion] (d2),
            (y1) -- [scalar, edge label'=$\Phi$] (y2)
        };
    \end{feynman}
    \fill[black] (0, 0) circle (2pt);
    \fill[black] (-1, 1) circle (2pt);
    \fill[black] (-1, -1) circle (2pt);
\end{tikzpicture}~~~~~
\begin{tikzpicture}
    \begin{feynman}
        \vertex (d1) at (-2.25, 1) {\(\large d_i\)};
        \vertex (d2) at (-2.25, -1) {\(\large d_j\)};
        \vertex (y1) at (-1, 1);
        \vertex (y1t) at (-1,1.3) {$\large y_{ik}$};
        \vertex (y2) at (-1,-1);
        \vertex (y1t) at (-1,-1.35) {$\large y_{jk}^{*}$};
        \vertex (a) at (0, 0);
        \vertex (b) at (1.5, 0);
        
        \diagram* {
            (y1) -- [scalar, edge label=$\Phi$] (a) -- [scalar, edge label=$\Phi$] (y2),
            (a) -- [boson, edge label=\( \gamma / g \)] (b),
             (a) -- [boson, edge label=\( \gamma / g \)] (b),
              (y1) -- [anti fermion] (d1),
            (y2) -- [fermion] (d2),
            (y1) -- [anti fermion, edge label = $d_k$] (y2)
        };
    \end{feynman}
    \fill[black] (0, 0) circle (2pt);
    \fill[black] (-1, 1) circle (2pt);
    \fill[black] (-1, -1) circle (2pt);
\end{tikzpicture}
\caption{\label{fig:dipoles} Diagrams for the contribution of either diquark to $d_i\to d_j \gamma$ or $d_i\to d_j g$.}
\end{center}
\end{figure}

\subsubsection*{Radiative meson decays}
The diquarks also produce a finite contribution to dipole operators through the diagrams in Fig.~\ref{fig:dipoles}. These enter radiative meson decays such as $B\to X_s \gamma$. These diagrams match to the SMEFT operators $O_{dB}$ and $O_{dG}$:
\begin{equation}
{\cal{L}}\supset C^{ij}_{dB} \left(\bar{q}_{i} \sigma^{\mu\nu}P_R d_{j} \right)H B_{\mu\nu} + C^{ij}_{dG} \left(\bar{q}_i \sigma^{\mu\nu}T^A P_R d_{j} \right)H G^A_{\mu\nu} + \text{h.c.}
\end{equation}
where $\sigma^{\mu\nu}=\frac{i}{2}[\gamma^\mu,\gamma^\nu]$, $q=(V^\dagger u_L, d_L)^T$ and $H=\frac{1}{\sqrt{2}}(0,v)^T$ with $v=246$ GeV. Our covariant derivative is defined $D_\mu=\partial_\mu + i e Q A^\mu + i g_s T^A G^A_\mu$. We then find for the triplet diquark:
\begin{equation}
\begin{split}
C_{dB}^{ij} &= -{g'\over  18\pi^2 M_{\Phi_{(3)}}^2}{m_i\over \sqrt{2}  v} y^{(3)*}_{ki} y^{(3)}_{kj} F \left({m_k^2 \over M_{\Phi_{(3)}}^2}\right), \\
C_{dG}^{ij} &= -{g_s\over  48\pi^2 M_{\Phi_{(3)}}^2}{m_i\over \sqrt{2}  v} y^{(3)*}_{ki} y^{(3)}_{kj} H\left({m_k^2 \over M_{\Phi_{(3)}}^2}\right), \\
\end{split}
\end{equation}
and for the sextet diquark:
\begin{equation}
\begin{split}
C_{dB}^{ij} &= -{g' \over  18\pi^2 M_{\Phi_{(6)}}^2}{m_i\over \sqrt{2}  v} y^{(6)*}_{ki} y^{(6)}_{kj} F\left({m_k^2 \over M_{\Phi_{(6)}}^2}\right), \\
C_{dG}^{ij} &= -{g_s \over  32\pi^2 M_{\Phi_{(6)}}^2}{m_i\over \sqrt{2}  v} y^{(6)*}_{ki} y^{(6)}_{kj} G\left({m_k^2 \over M_{\Phi_{(6)}}^2}\right), \\
\end{split}
\end{equation}
with 
\begin{equation}
\begin{split}
F(x)& = {4-9x+5x^3+6x(1-2x)\log x \over 4(1-x)^4} ,\\
H(x)& = {1 + 9x -9x^2-x^3+6x(1+x)\log x \over (1-x)^4}, \\
G(x)& = {1 - 11 x + 7 x^2 +3 x^3-2x(1+5x)\log x \over  (1-x)^4} .
\end{split}
\end{equation}
To our knowledge, the sextet calculations have not previously appeared in the literature.
In the summary plots Fig.~\ref{fig:collidervsflavour3} and Fig.~\ref{fig:collidervsflavour6}, the resulting constraints from radiative decays are included in the blue regions found using the \texttt{smelli} quark likelihood~\cite{Aebischer:2018iyb,Aebischer:2018bkb,Straub:2018kue}.

\subsubsection*{$\epsilon^\prime/\epsilon$ ratio}
\begin{figure}[p]
\centering
\includegraphics[width=0.45\textwidth]{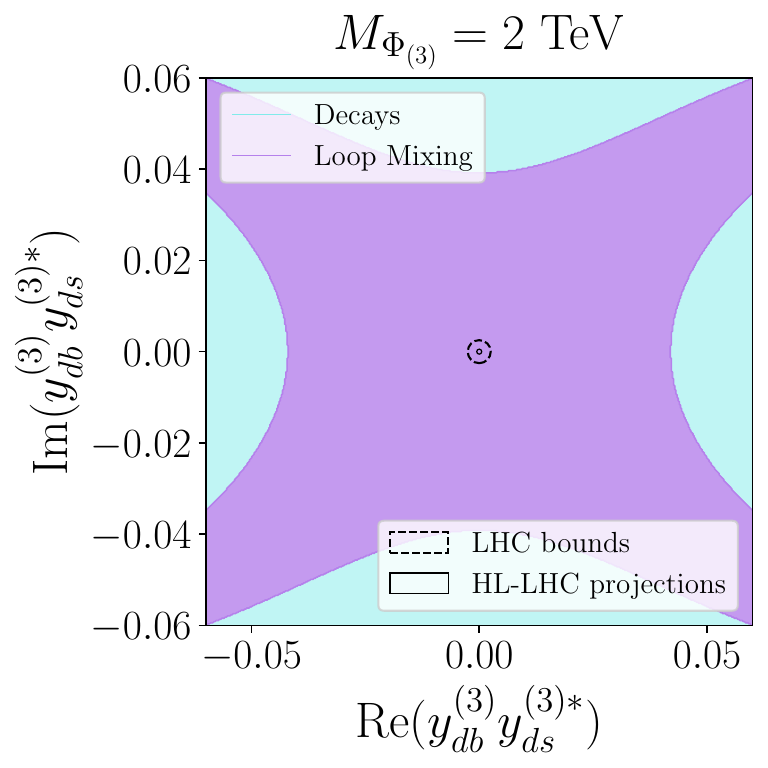}~~~
\includegraphics[width=0.45\textwidth]{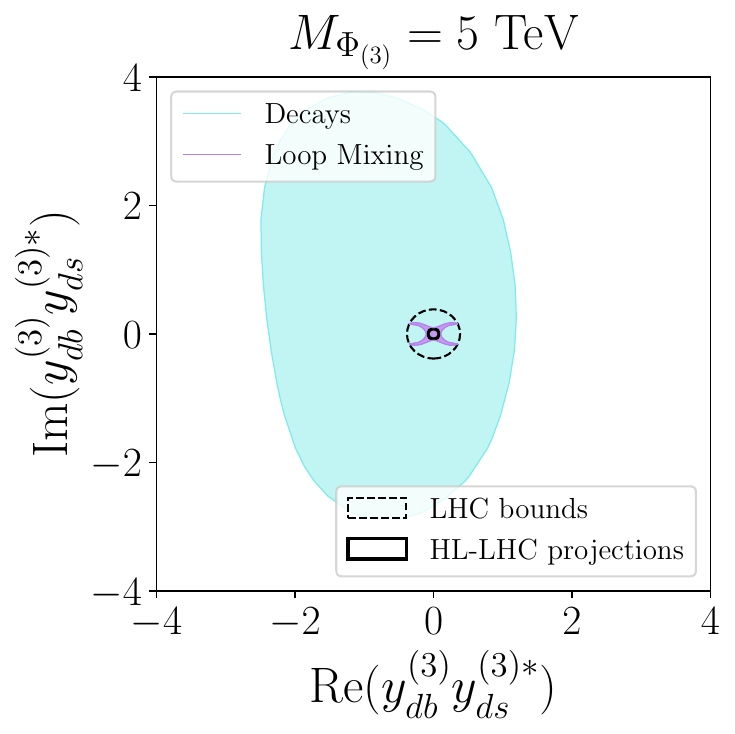}\\
\vspace*{5mm}
\includegraphics[width=0.45\textwidth]{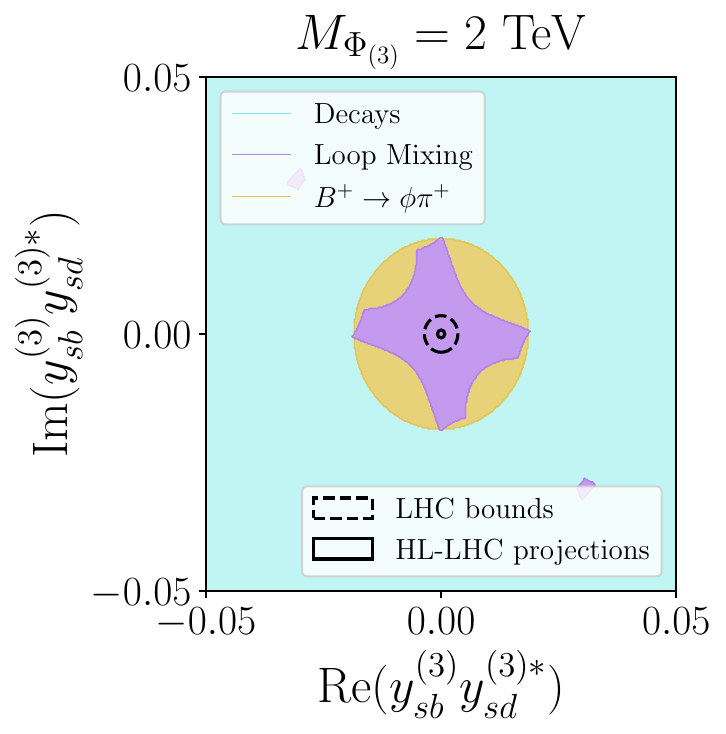}~~~
\includegraphics[width=0.45\textwidth]{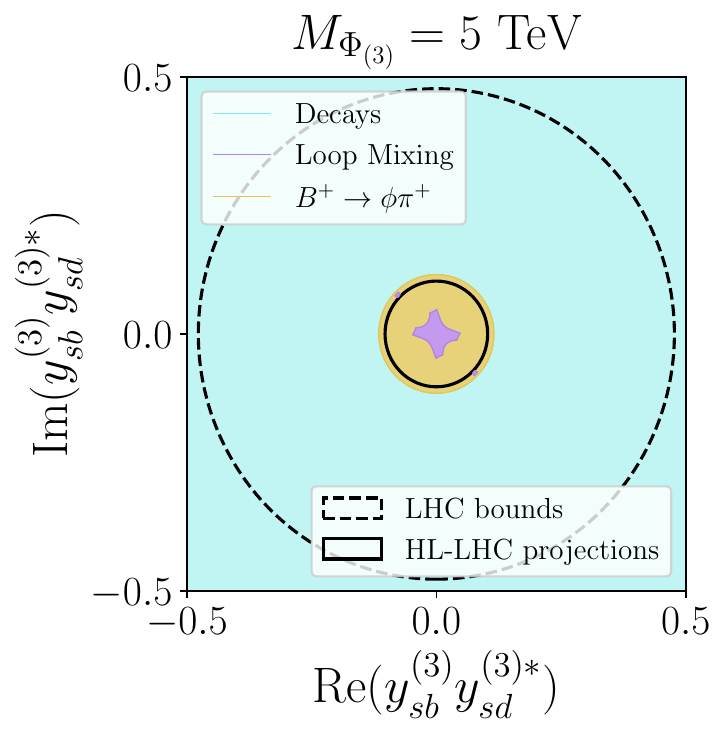} \\
\vspace*{5mm}
\includegraphics[width=0.45\textwidth]{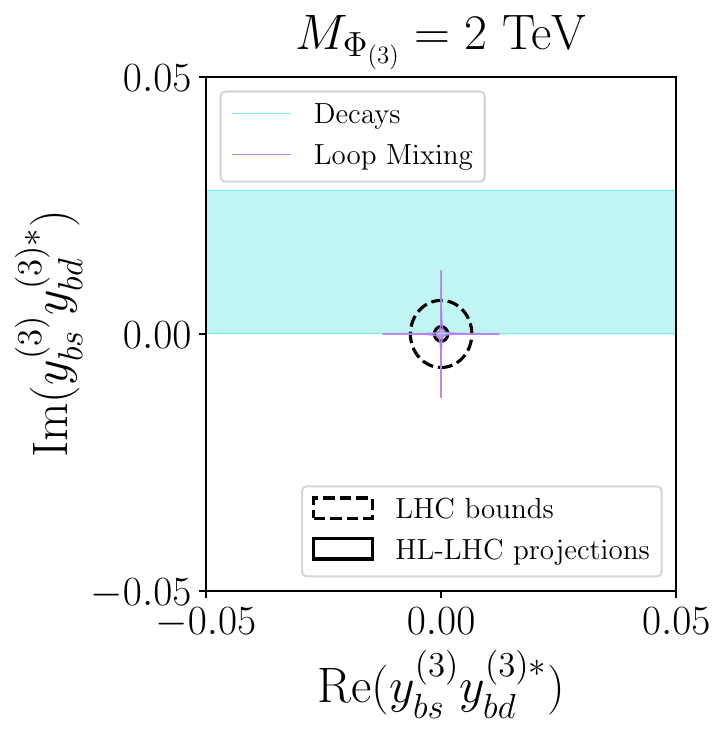}~~~
\includegraphics[width=0.45\textwidth]{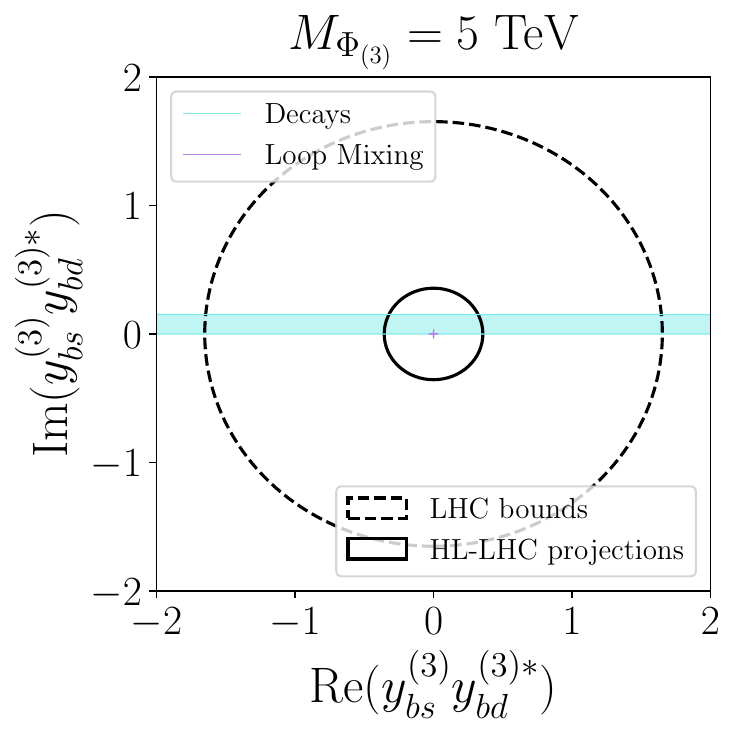}
\caption{Allowed regions (95\% CL) in the real and imaginary parts of different products of couplings for the triplet diquark. Left column: mass of 2 TeV, right column: mass of 5 TeV. In each case, all other couplings are set to zero.}
\label{fig:collidervsflavour3}
\end{figure}

\begin{figure}[p]
\centering
\includegraphics*[width=0.45\textwidth]{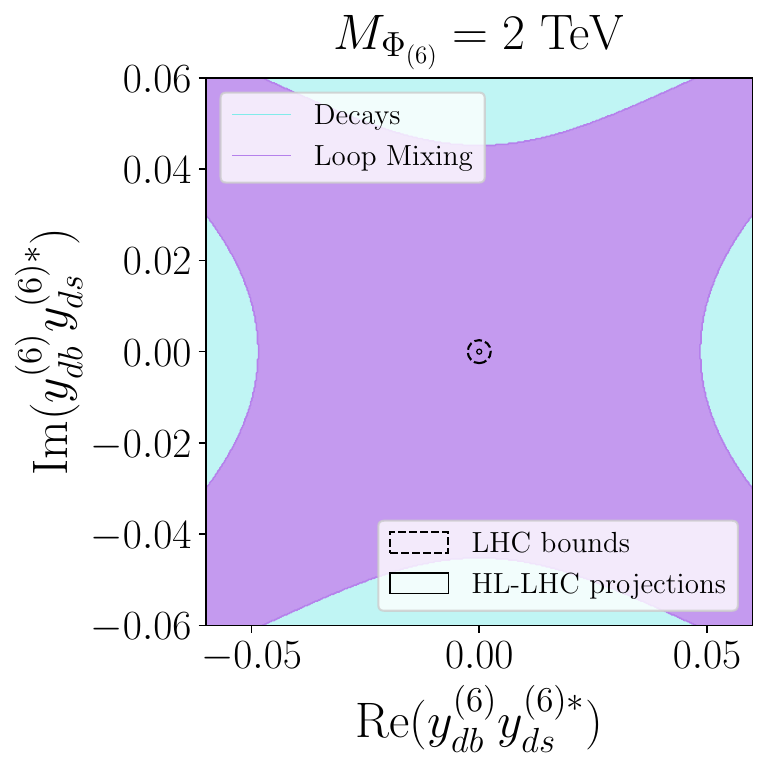}~~~
\includegraphics[width=0.45\textwidth]{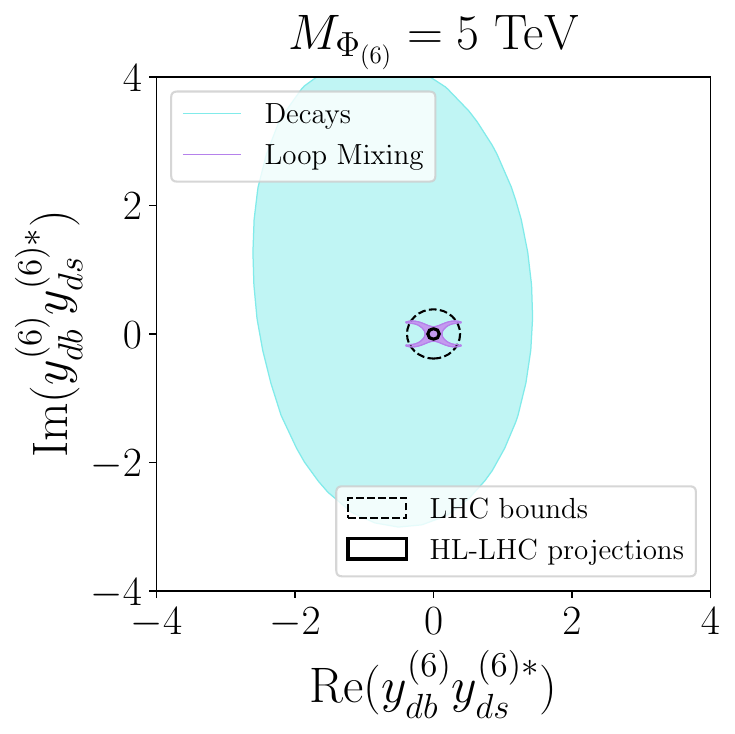} \\
\vspace*{5mm}
\includegraphics[width=0.45\textwidth]{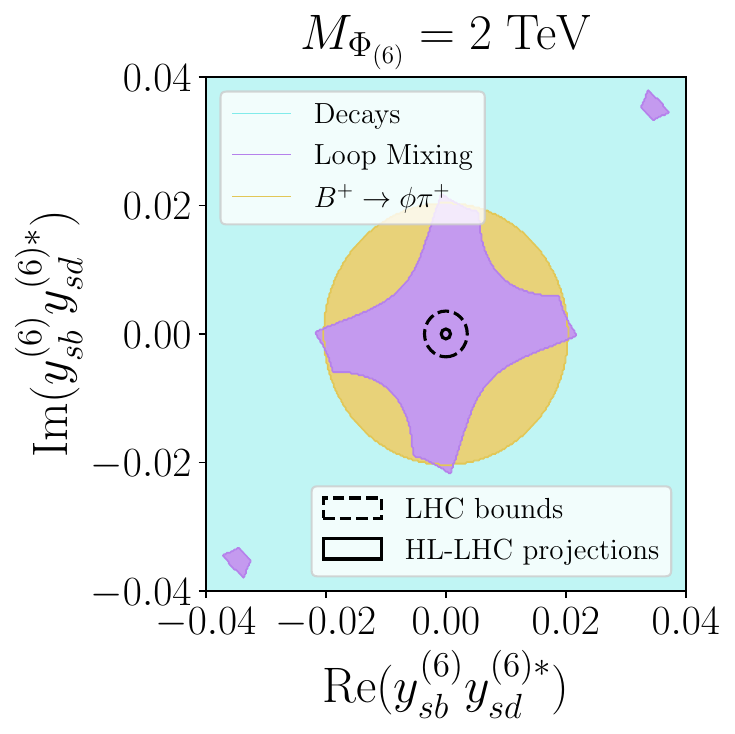}~~~
\includegraphics[width=0.45\textwidth]{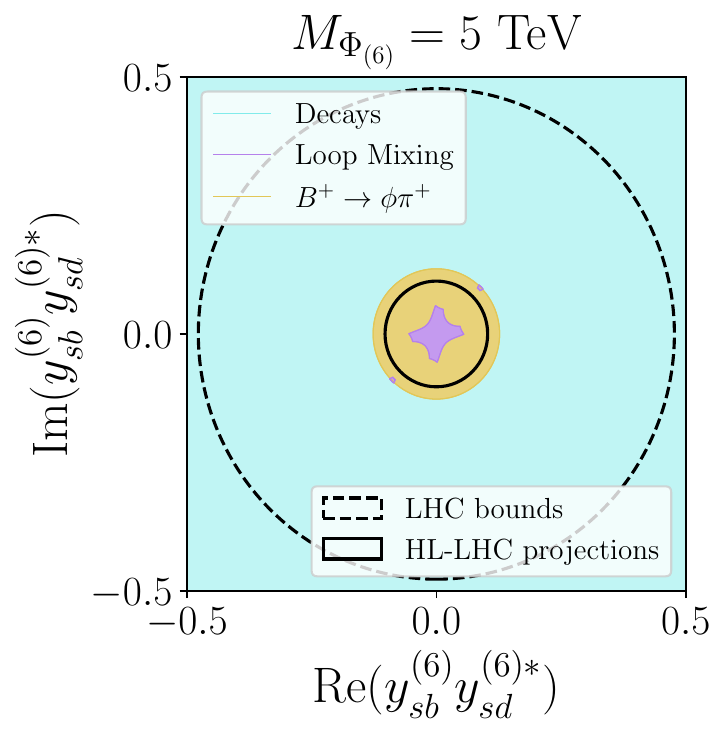} \\
\vspace*{5mm}
\includegraphics[width=0.45\textwidth]{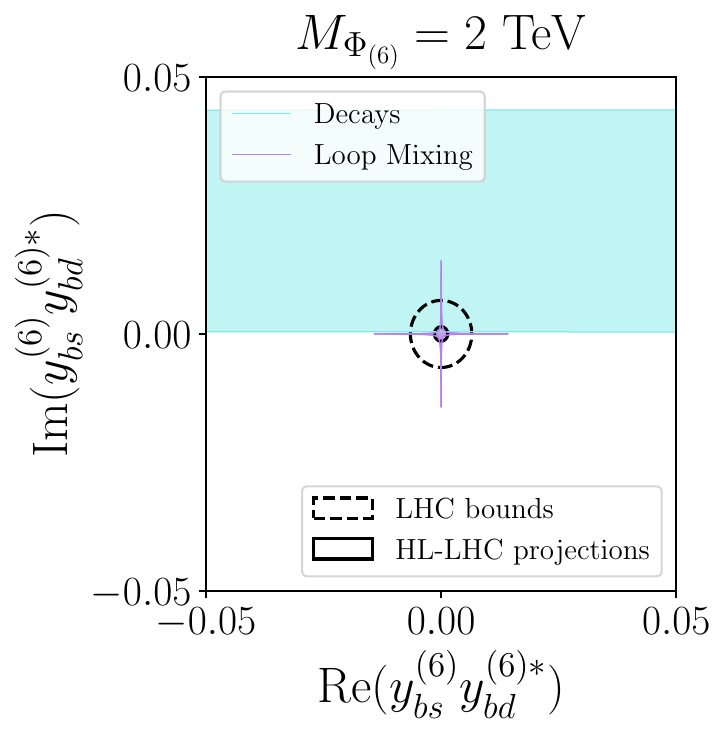}~~~
\includegraphics[width=0.45\textwidth]{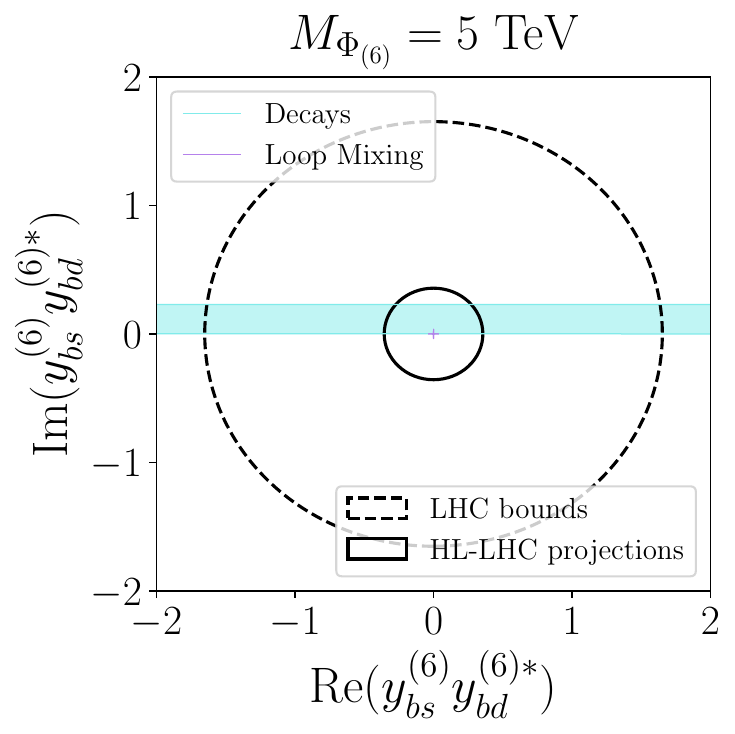}
\caption{Allowed regions (95\% CL) in the real and imaginary parts of different products of couplings for the sextet diquark. Left column: mass of 2 TeV, right column: mass of 5 TeV. In each case, all other couplings are set to zero.}
\label{fig:collidervsflavour6}
\end{figure}

The $\epsilon^\prime/\epsilon$ ratio measures the size of direct CP violation in $K\to \pi\pi$ decays relative to CP violation in neutral Kaon mixing.
A master formula for the effects of new physics in this observable has been presented in Refs~\cite{Aebischer:2018csl,Aebischer:2018quc,Aebischer:2021hws}. The diquarks can contribute to $\epsilon^\prime/\epsilon$ in three different ways: through four-quark operators with $\Delta S=1$, through renormalisation group mixing into gauge boson vertex corrections (as in Fig.~\ref{fig:RGEZ} but without a lepton current), and through loop contributions to the chromomagnetic dipole operators $s\to d g$ (as calculated in the previous section).
We include all of these contributions to this observable in constructing the blue region in our plots in Figs~\ref{fig:collidervsflavour3} and~\ref{fig:collidervsflavour6} using the \texttt{smelli} quark likelihood~\cite{Aebischer:2018iyb,Aebischer:2018bkb,Straub:2018kue}.

\section{Flavour meets Collider}
\label{sec:flavourcoll}
In Figs~\ref{fig:collidervsflavour3} and \ref{fig:collidervsflavour6}, we superimpose collider and flavour constraints on the couplings of the diquarks, at both a mass of 2 TeV and a mass of 5 TeV. In each plot, only the couplings on the axes are non-zero.

The flavour constraints are separated into two pieces: meson mixing at loop level and `decays', which includes (semi)leptonic and radiative meson decays and $\epsilon^\prime/\epsilon$. For the $y_{sb}y_{sd}^*$ coupling product, there is additionally a constraint from the branching ratio of the decay $B^+\to \pi^+\phi$. The calculations that go into each of these constraints are described in the previous Section~\ref{sec:offdiagflavour}. The blue `decays' region is calculated using the \texttt{smelli} quark likelihood~\cite{Aebischer:2018iyb,Aebischer:2018bkb,Straub:2018kue}. The particular flavour observables which produce the constraints in the various plots are generally determined by the couplings; specifically, the uppermost $y_{db}y_{ds}^*$ plots display constraints from $B_s$ mixing and meson decays involving $b\to s$ transitions, the middle $y_{sb}y_{sd}^*$ plots display constraints from $B_d$ mixing and meson decays involving $b\to d$ transitions, while the lower $y_{sb}y_{sd}^*$ plots display constraints from $K$ mixing and meson decays involving $s\to d$ transitions. For the triplet diquark, the products of couplings on the axes in Fig.~\ref{fig:collidervsflavour3} are then the only ones involved in the relevant flavour observables. By contrast, to obtain the constraints in Fig.~\ref{fig:collidervsflavour6} for the sextet diquark, we have explicitly set its diagonal couplings $y^{(6)}_{ii}$ to zero. This can be taken as a reasonable phenomenological assumption, since (products of) the diagonal couplings are so strongly constrained in general by meson mixing, as discussed in Sec.~\ref{sec:diagflavour}. With this assumption, it can be seen that the flavour constraints are comparable between the triplet and the sextet. This can be understood from the fact that the sextet and triplet result in generally very similar expressions for the flavour observables and bounds in Sec.~\ref{sec:offdiagflavour}. The blue `decays' regions are slightly different, due mainly to the fact that some of the decays are generated via RGEs from the four-quark operators, which do not receive equal matching contributions from the two diquarks,~Eqs~\eqref{eq:sextetmatching} and \eqref{eq:tripletmatching}.

The collider constraints shown in Figs~\ref{fig:collidervsflavour3} and \ref{fig:collidervsflavour6} are obtained by setting the relevant couplings to equal values, and choosing the remaining couplings to be zero. The bounds are derived from a two-sigma limit of the bump hunt at a given mass over the continuum background as detailed in Sec.~\ref{sec:bump}. For our conventions, upon specifying the couplings given in these figures, this amounts to selecting combinations of different partonic subprocesses. At a given mass and coupling, the cross section then directly reflects the parton luminosity dependence of single resonance triplet or sextet production. This explains the reduced sensitivity of the collider constraints when moving, e.g., down the columns of Fig.~\ref{fig:collidervsflavour3}. The down-type (valence quark) luminosities are larger than the $s$ luminosities, which, in turn, are more sizeable than the bottom sea quark densities. Hence the cross section constraints become increasingly loose when moving from top to bottom in the figure.

The plots in Figs~\ref{fig:collidervsflavour3} and \ref{fig:collidervsflavour6} demonstrate the complementarity of collider and flavour observables to narrow in on the diquark parameter space. At masses of 2 TeV, collider constraints are generally stronger than flavour constraints, but at masses of 5 TeV, flavour constraints will typically dominate. The different scaling of their sensitivity with mass therefore allows these datasets to probe different areas of diquark parameter space. Both collider and flavour measurements have the potential to discover down-type diquarks, and both are needed to explore the full parameter space and coupling structure. In combination, they rule out a down-type diquark to multi-TeV masses, as long as at least two of its couplings are order one. Importantly, there exist parameter regions that are not yet probed at the LHC, for which flavour physics enables the discrimination between the triplet and the sextet, while their single-resonance production is largely identical. This means that if a discovery is made at the LHC that points to strongly interacting new physics, including the flavour data set can break the parameter degeneracy of the down-type diquark interpretation. This also shows that flavour-collider complementarity will become increasingly important when moving towards the high-luminosity phase of the LHC.

\section{Summary and Conclusions}
\label{sec:conc}
Almost a decade and a half into the successful operation of the LHC, an important effort is underway to take stock of the remaining hiding places for new physics. The model-independent framework of the SMEFT can help to organise this effort, providing an efficient method of surveying constraints on the full BSM landscape. But in certain parts of this landscape, the operator bounds are so weak, and the number of tree-level UV completions so few, that it is worthwhile to instead study these UV completions directly. In this article, we have focussed on one such case: two scalar diquarks which, by virtue of their SM gauge charges, can only match at tree level to the SMEFT operator $O_{dd}$ containing four right-handed down-type quarks.

We have studied the signals expected from these states in LHC collisions, in Higgs physics, and in flavour observables. At the LHC, the diquarks can be singly-produced through their Yukawa couplings to quarks or pair-produced via their QCD interactions. We show that current dijet bump-hunt searches constrain these diquarks to multi-TeV scales for $O(1)$ Yukawa couplings and that these bounds will improve significantly by the end of HL-LHC. The leading LHC bump-hunt bounds are generally from single production, so they can vary depending on the size and flavour structure of the diquark couplings due to differing parton luminosities of the different quarks involved in the exotics' production. But even for extremely small couplings, pair-production bounds cannot be evaded and constrain the diquarks to be heavier than a TeV.

Flavour observables, by contrast, can test the models to hundreds of TeV scales, but their sensitivity is entirely dependent on the size and flavour structure of the diquark couplings and can be drastically reduced under motivated flavour assumptions, as we demonstrate via a partial compositeness paradigm. The sensitivity of meson mixing observables also depends strongly on \emph{which} down-type diquark is present; the colour sextet can contribute at tree level to these observables, while the triplet's leading contributions are only at one loop. We have catalogued all important flavour observables that can constrain these diquarks, including loop-level constraints, which are sensitive to the off-diagonal couplings of both~states. 

Due to their different sensitivity dependence on the coupling and mass of the diquarks, flavour and collider information can combine powerfully for these models. Diquarks with masses of several TeV can often evade one or the other, depending on their particular couplings, but are unlikely to evade both. We illustrate this by showing constraints in several different planes in parameter space. This complementarity also enables characterisation of the model in case of a discovery; different coupling combinations can produce degenerate signals in a dijet bump hunt, but very different predictions in flavour-changing processes. 

In summary, down-type scalar diquarks occupy a special position in BSM parameter space. They match at tree level only onto a single (poorly-constrained) SMEFT operator, and as $SU(2)_L$ singlets their loop effects in electroweak precision observables are very small. On the other hand, their QCD charges and flavoured interactions predict a range of striking signals at colliders and in flavour observables. By focussing on these simple UV completions, we cut into an important part of the parameter space for new interactions involving down-type quarks.

\subsection*{Acknowledgements}
We thank Joe Davighi, Matthew Kirk, Peter Stangl, and Gilberto Tetlalmatzi-Xolocotzi for helpful discussions.
C.E. and S.R. are supported by the UK Science and Technology Facilities Council (STFC) under grant ST/X000605/1. C.E. acknowledges further support from the Leverhulme Trust under Research Project Grant RPG-2021-031 and Research Fellowship RF-2024-300$\backslash$9. C.E. is supported by the Institute for Particle Physics Phenomenology Associateship Scheme. S.R. is supported by UKRI Stephen Hawking Fellowship EP/W005433/1. W.N. is funded by a University of Glasgow College of Science of Engineering fellowship.

\appendix
\section{Details on composite scenarios}
\label{sec:cosetapp}
In this appendix we provide a few more details on how the symmetry breaking occurs to obtain the diquarks as composite pseudo-Nambu Goldstone bosons (PNGBs), as proposed in Sec.~\ref{sec:comphiggs}. We take as a starting point that the Higgs is obtained as a PNGB from the $SO(5)/SU(2)_L\times SU(2)_R$ part of the cosets in Eqns.~\eqref{eq:cosetsextet} and \eqref{eq:cosettriplet} in an analogous way to the well-known minimal composite Higgs model \cite{Agashe:2004rs}. So here we focus on the extension of the coset structure needed to additionally produce the diquarks. We also provide possible charge assignments of the embeddings of the fundamental fermions under the unbroken subgroup.
\subsection{$Sp(6)/SU(3)\times U(1)$}
\label{sec:sp6}
\subsubsection*{Coset structure}
$Sp(2N)$ is defined as the group of unitary matrices $U$ with unit determinant for which
$U\Omega U^T=\Omega$, where $\Omega$ is the $2N\times 2N$ matrix
\begin{equation}
\Omega=\begin{pmatrix}
0 & I_N \\ -I_N & 0
\end{pmatrix},
\end{equation}
and $I_N$ is the $N\times N$ unit matrix. From this definition, the generators $K$ of $Sp(2N)$ (with $U\equiv 1+iK$) are the set of $2N^2+N$ Hermitian matrices 
\begin{equation}
K=\begin{pmatrix}
A & B \\ B^\dagger &-A
\end{pmatrix} \text{with $N\times N$ matrices $A$, $B$ such that }B=B^T, ~A=A^\dagger.
\end{equation}
A basis of the 21 generators for $Sp(6)$ can hence be written (where $\lambda_i$ are the usual Gell-Mann matrices which, along with the identity matrix $I_3$, form a basis for the $3\times 3$ Hermitian matrices $A$):
\begin{align}
K_{i=1,...,8}= \begin{pmatrix}
\lambda_i & 0 \\
0 & -\lambda_i
\end{pmatrix},~~ K_9=\begin{pmatrix}
I_3 & 0 \\
0 & -I_3
\end{pmatrix}, ~~K_{9+j}= \begin{pmatrix}
0 & S_j \\
S^\dagger_j & 0
\end{pmatrix},
\end{align}
where $S_j=S_j^T$ ($j=1,...,12$) are a basis of the complex symmetric $3\times 3$ matrices. The breaking of the $Sp(6)$ group down to $SU(3)\times U(1)$ can be achieved by a field in the adjoint representation of $Sp(6)$ which takes a vev of the form $\langle \Phi \rangle \propto K_9$.
The unbroken generators $K_i$ for $i=1,...,9$ satisfy $[\langle \Phi \rangle, K_i]=0$ and correspond to the generators of $SU(3)\times U(1)$, while the remaining broken generators satisfy $\{\langle \Phi \rangle, K_j\}=0$.

\subsubsection*{Couplings to fermions}
\label{sec:compfermionsextet}
In the partial compositeness paradigm, the SM fermions couple linearly to a strong sector operator, and the resultant mixing is the origin of the fermion masses as well as the couplings to the Higgs boson and to the diquark. To write down these couplings, the fermions must be embedded into representations of $\mathcal{H}\sim SU(3)_c\times U(1)_s \times SU(2)_L\times SU(2)_R \times U(1)_X$. The quarks can be embedded in the $\mathbf{6}$ representation of $Sp(6)$, which decomposes as $\mathbf{3}_1 \oplus \bar{\mathbf{3}}_{-1}$ under $SU(3)_c \times U(1)_s$.\footnote{For branching rules, see e.g.~\cite{Feger:2019tvk}.} On the electroweak side of the coset, the quark doublet should be within the $(\mathbf{2},\mathbf{2})$ representation of $SU(2)_L\times SU(2)_R$ in order to avoid large corrections to the $Zb_L\bar{b}_L$ coupling~\cite{Agashe:2006at}, which embeds within a $\mathbf{5}$ of $SO(5)$, along with a singlet of both $SU(2)_L$ and $SU(2)_R$, in which one of $d_R$ or $u_R$ can be embedded. The other one of $d_R$ or $u_R$ must instead be a triplet of $SU(2)_R$ (such that $d_R$ and $u_R$ end up with different hypercharges, but all Yukawa couplings are still gauge invariant), which can be embedded within the $\mathbf{10}$ irrep of $SO(5)$. Overall, using $T_Y=-\frac{1}{3}T_s +T_{3R}+T_X$, one viable embedding of the SM fermions within $SU(3)_c\times U(1)_s \times SU(2)_L\times SU(2)_R \times U(1)_X$ is as part of the representations:
\begin{align}
Q&\in (\mathbf{3},1,\mathbf{2},\mathbf{2},0), ~u\in (\mathbf{3},1,\mathbf{1},\mathbf{3},0),~d\in (\mathbf{3},1,\mathbf{1},\mathbf{1},0),\nonumber \\
L&\in (\mathbf{1},0,\mathbf{2},\mathbf{2},-1), ~e\in (\mathbf{1},0,\mathbf{1},\mathbf{1},-1).
\end{align}
Note that the Higgs and the diquark are uncharged under $U(1)_X$.

\subsection{$SO(6)/SU(3)\times U(1)$}

\subsubsection*{Coset structure}
We will use the fact that $SO(6)$ is isomorphic to $SU(4)$, and work within the representations of $SU(4)$ here. A basis of the 15 generators of $SU(4)$ can be written
\begin{align}
&T_{i=1,...,8}=\begin{pmatrix}
 & & & 0\\
 & \lambda_i & & 0 \\
 & & & 0\\
 0 & 0 & 0 & 0
\end{pmatrix}, ~T_9=\begin{pmatrix}
0 &0 & 0& 1\\
0 & 0 & 0 & 0 \\
0  &0  & 0& 0\\
 1 & 0 & 0 & 0
\end{pmatrix}, ~T_{10}=\begin{pmatrix}
0 &0 & 0& -i\\
0 & 0 & 0 & 0 \\
0  &0  & 0& 0\\
i & 0 & 0 & 0
\end{pmatrix}, ~T_{11}=\begin{pmatrix}
0 &0 & 0& 0\\
0 & 0 & 0 & 1 \\
0  &0  & 0& 0\\
0 & 1 & 0 & 0
\end{pmatrix},\nonumber \\
&T_{12}=\begin{pmatrix}
0 &0 & 0& 0\\
0 & 0 & 0 & -i \\
0  &0  & 0& 0\\
0 & i & 0 & 0
\end{pmatrix}, ~T_{13}=\begin{pmatrix}
0 &0 & 0& 0\\
0 & 0 & 0 & 0 \\
0  &0  & 0& 1\\
0 & 0 & 1 & 0
\end{pmatrix},~T_{14}=\begin{pmatrix}
0 &0 & 0& 0\\
0 & 0 & 0 & 0 \\
0  &0  & 0& -i\\
0 & 0 & i & 0
\end{pmatrix}, ~T_{15}=\frac{1}{\sqrt{6}}\begin{pmatrix}
1 &0 & 0& 0\\
0 & 1 & 0 & 0 \\
0  &0  & 1& 0\\
0 & 0 & 0 & -3
\end{pmatrix},
\end{align}
where $\lambda_i$ in $T_{i=1,...,8}$ should be taken to mean that the upper left $3\times 3$ block of these generators are the Gell-Mann matrices. The breaking of the $SU(4)$ group down to $SU(3)\times U(1)$ can be achieved by a field in the adjoint representation of $SU(4)$ which takes a vev of the form $\langle \Phi \rangle \propto T_{15}$.
The unbroken generators $T_i$ for $i=1,...,8$  and $i=15$ satisfy $[\langle \Phi \rangle, T_i]=0$ and correspond to the generators of $SU(3)\times U(1)$, while the remaining broken generators have $[\langle \Phi \rangle, T_j]\neq 0$.

\subsubsection*{Couplings to fermions}
To write down the couplings of fermions to the diquark and the Higgs, the fermions must be embedded into representations of $\mathcal{H}\sim SU(3)_c\times U(1)_t \times SU(2)_L\times SU(2)_R \times U(1)_X$. For consistency we continue to work within the representations of $SU(4)$ when considering the broken group $\mathcal{G}$. The quarks can be embedded in the $\mathbf{6}$ representation of $SU(4)$, which decomposes as $\mathbf{3}_{-2} \oplus \bar{\mathbf{3}}_{2}$ under $SU(3)_c \times U(1)_t$. On the electroweak side of the coset, the charges under $SO(5)$ can be taken to be the same as in Sec.~\ref{sec:sp6} above. Overall, using $T_Y=\frac{1}{6}T_t +T_{3R}+T_X$, one possible viable embedding of the SM fermions within $SU(3)_c\times U(1)_t \times SU(2)_L\times SU(2)_R \times U(1)_X$ is as part of the representations:
\begin{align}
Q&\in (\mathbf{3},-2,\mathbf{2},\mathbf{2},0), ~u\in (\mathbf{3},-2,\mathbf{1},\mathbf{3},0),~~d\in (\mathbf{3},-2,\mathbf{1},\mathbf{1},0),\nonumber \\
L&\in (\mathbf{1},0,\mathbf{2},\mathbf{2},-1), ~e\in (\mathbf{1},0,\mathbf{1},\mathbf{1},-1).
\end{align}

\bibliographystyle{JHEP}
\bibliography{references}

\end{document}